\documentclass{article}
\usepackage[utf8]{inputenc}
\usepackage{amsmath}
\usepackage{graphicx}
\usepackage{algorithm}
\usepackage{algpseudocode}
\usepackage[margin=3 cm]{geometry}
\usepackage{caption}
\usepackage{subcaption}
\usepackage{siunitx}
\usepackage{authblk}
\title{A Delaunay Refinement Algorithm for the Particle Finite Element Method applied to Free Surface Flows}
\author[1,*]{Thomas Leyssens}
\author[1]{Michel Henry}
\author[1]{Jonathan Lambrechts}
\author[1]{Jean-François Remacle}
\affil[1]{Institute of Mechanics, Materials and Civil Engineering, Université catholique de Louvain, Louvain-la-Neuve, Belgium}
\affil[*]{Correspondence: Thomas Leyssens, thomas.leyssens@uclouvain.be}
\date{}

\begin{document}

\maketitle
\section*{Abstract}

This paper proposes two contributions to the calculation of free-surface flows using the particle finite element method (PFEM).

The PFEM is based upon a Lagrangian approach: a set of particles defines the fluid and each particle is associated with a velocity vector. 
Then, unlike a pure Lagrangian method, all the particles are connected by a triangular mesh. 
The difficulty lies in locating the free surface from this mesh. 
It is a matter of deciding which of the elements in the mesh are part of the fluid domain, and to define a boundary -- the free surface.
Then, the incompressible Navier-Stokes equations are solved on the fluid domain and the particle position is updated using the velocity vector from the finite element solver. 

\emph{Our first contribution is to propose an approach to adapt the mesh with theoretical guarantees of quality}: 
the mesh generation community has acquired a lot of experience and understanding about mesh adaptation approaches with guarantees of quality on the final mesh. 
The approach we use here is based on a \emph{Delaunay refinement strategy}, allowing to insert and remove nodes while gradually improving mesh quality. 
We show that what is proposed allows to create stable and smooth free surface geometries.

One characteristic of the PFEM is that only one fluid domain is modelled, even if its shape and topology change. 
It is nevertheless necessary to apply conditions on the domain boundaries. 
When a boundary is a free surface, the flow on the other side is not modelled, it is represented by an external pressure.  
On the external free surface boundary, atmospheric pressure can be imposed. 
Nevertheless, there may be internal free surfaces: 
the fluid can fully encapsulate cavities to form bubbles. 
The pressure required to maintain the volume of those bubbles is \emph{a priori} unknown.
For example, the atmospheric pressure would not be sufficient to prevent the bubbles from deflating and eventually disappearing. 
\emph{Our second contribution is to propose a multi-point constraint approach to enforce global incompressibility of those empty bubbles.}
We show that this approach allows to accurately model bubbly flows that involve two fluids with large density differences, for instance water and air, while only modelling the heavier fluid.
\section{Introduction}

A considerable number of flows involve \emph{free surfaces} that may evolve in complex geometries. 
In many cases, these flows are highly transient. 
A wide range of applications of free surface flows can be found in various fields of engineering: 
hydraulics (flows over weirs, around bridge piers, dam breaks, wetting and drying), 
naval engineering (flows around ship hulls), 
additive manufacturing (a melt pool is produced as a result of laser radiation on the surface of the powder), combustion, etc. 

The main difficulty in numerically computing free surface flows is directly linked to modelling the geometry of the free surface. 
Indeed, the free surface evolves in time, sometimes rapidly, changing its scales, its geometry and  its topology.
The computational fluid dynamics community has long been interested in modelling strongly deforming free surface flows. 
Two classes of methods have emerged for solving them: the evolving free surface can be numerically \emph{tracked} or \emph{captured}.  

Interface capturing methods represent the interface on a fixed mesh and let it evolve by means of an advection equation.
By contrast, in interface tracking approaches, the mesh is conforming to the interface at all times. 
This is of course a challenge: consider, for example, a seemingly simple fluid mechanics problem of a container partially filled with water. 
When this container is shaken, the dynamics of the water become relatively complex: 
interactions with the container walls, mixing of the water with the surrounding air, droplets splashing, etc. 

The particle finite element method, or PFEM, is a tracking approach designed to tackle this complexity. 
Initially developed for fluid flows \cite{pfemInitial}, the PFEM has been successfully applied to complex industrial applications 
such as fluid structure interactions \cite{cerquagliaFSI}, landslides \cite{landslides}, or even solid mechanics problems such as metal cutting \cite{metal}.

The PFEM starts from a Lagrangian point of view:
the fluid domain is represented by a set of particles that follow the motion of the fluid.
These particles carry all the relevant fields (velocity, pressure, temperature, etc.), but they have no prescribed mass. 
Lagrangian methods are typically separated into two different categories: meshless methods and mesh-based methods. 
The smoothed particle hydrodynamics (SPH) \cite{sph} and material point method (MPM) \cite{mpm} belong to the meshless group. 
These methods have the advantage of not requiring any remeshing.
However, the absence of a mesh leads to additional difficulties.
Computing gradients of fields can be tricky and properly imposing boundary conditions is challenging.
In mesh-based Lagrangian methods, particles are connected through a mesh that is used to 
solve the equations describing the evolution of the continuum (e.g. with finite elements). If the connectivity of the mesh is fixed,
large deformations of the front are not always permitted: arbitrary Lagrangian-Eulerian (ALE) \cite{ale} 
methods are extremely accurate methods with great properties of conservation but they are restricted to
small deformations of the free surface and have difficulties with topological changes of the fluid domain.

\begin{figure}
    \centering
    \includegraphics[width=.9\textwidth]{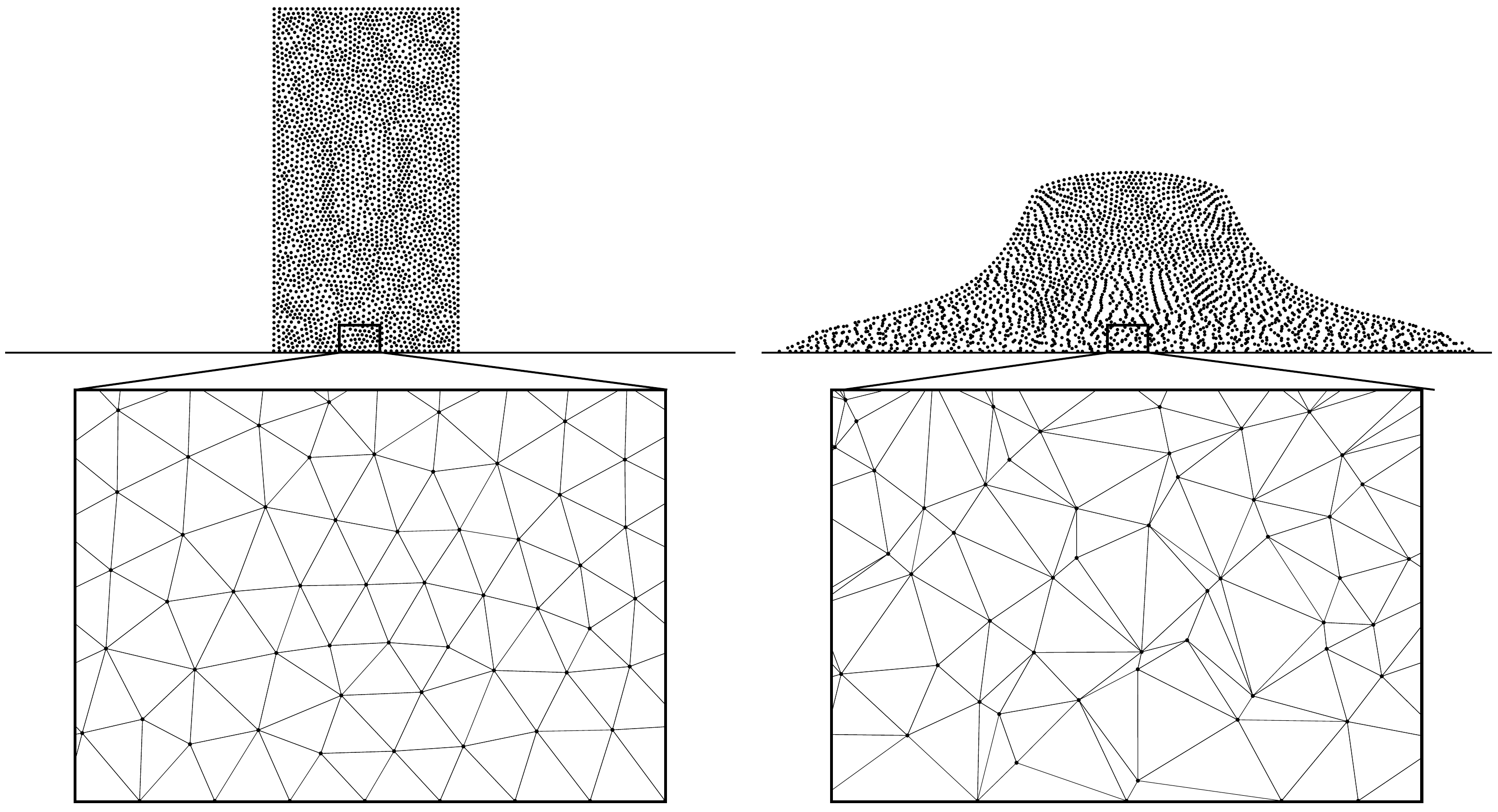}
    \caption{A typical PFEM simulation of a water column collapsing. 
            After a number of time steps, the initially well-distributed particles move according to the velocity field, 
            resulting in a disordered state.
            Without any adaptation, the mesh quality suffers greatly from this motion of the particles, eventually leading to poor quality simulations.
            }\label{particlesMesh}
\end{figure}

Though the PFEM \cite{cremonesiPFEM,onatePFEM} is mesh-based, this method actually assumes that computing the Delaunay triangulation/tetrahedrization of a large set of points is a task that can be carried out in an efficient manner. 
In other words, recomputing the triangulation at every time step is cheaper than solving the Navier-Stokes equations \cite{celestin}. 
One remaining difficulty is to determine the boundary of the fluid domain -- the free surface. 
This challenge is linked to a well-known problem of computational geometry: 
surface reconstruction from point clouds. 
We need an indicator that defines the boundary of the domain.
A Delaunay triangulation covers exactly the convex hull of the set of points: this is the simplest indicator. 
More complex indicators based on $\alpha$-shapes have been proposed in the PFEM. 
Using $\alpha$-shapes roughly means eliminating poor-quality elements from the triangulation (too large or too distorted). 
If we consider a set of points that is "ordered" (like in a finite element mesh), the $\alpha$-shape approach works pretty well. 
Let's now imagine that our set of points is initially well distributed.  
After a few time steps, the points will evolve with the fluid velocity into a more disordered state, 
with some regions highly concentrated in particles, and others suffering from low particles density. 
This leads to a significant degradation in mesh quality -- even inside the fluid domain. 
Figure \ref{particlesMesh} illustrates this phenomenon.

The fact that the mesh is of poor quality has two consequences. 
First, as with any other finite element simulation, solving the equations on a quality mesh is important for obtaining a good accuracy. 
Second, even more crucially, in the context of the PFEM, the detection of the fluid domain actually depends on mesh quality. 
Hence, degrading the mesh has the well-known consequence of causing abrupt, unphysical variations in the shape of the fluid domain. 
Solutions have been proposed \cite{addNodes} in the literature for adapting the mesh during simulation: 
adding nodes when the local density is too low, or removing nodes when they are too close. 

In this paper, our first contribution is to \emph{propose an approach to adapt the mesh with theoretical guarantees of quality}: 
the mesh generation community has acquired a lot of experience and understanding about mesh adaptation approaches with quality guarantees on the final mesh. 
The approach we use here is based on a \emph{Delaunay refinement strategy}, more specificaly Chew's first algorithm \cite{chew}, 
allowing to insert and remove nodes while gradually improving mesh quality. 
The philosophy consists of enforcing a high quality Delaunay mesh at all times, and refining it based on a user-defined size field.
This greatly improves the robustness of the method, as it removes the arbitrariness in the detection of the free surface. 
Moreover, the use of a size field is very useful since it allows greater accuracy in regions of interest through mesh refinement, 
and saves computational costs thanks to coarsening. 

One characteristic of the PFEM is that only one fluid domain is modelled, even if its shape and topology change. 
It is nevertheless necessary to apply conditions on the domain boundaries. 
When a boundary is a free surface, the flow on the other side is not modelled, it is represented by an external pressure.  
On the external free surface boundary, atmospheric pressure can be imposed. 
Nevertheless, there may be internal free surfaces: 
the fluid can fully encapsulate cavities to form bubbles. 
The pressure required to maintain the volume of those bubbles is \emph{a priori} unknown.
For example, the atmospheric pressure would not be sufficient to prevent the bubbles from deflating and eventually disappearing. 
\emph{Our second contribution is to propose a multi-point constraint approach to enforce global incompressibility of those empty bubbles.}
We show that this approach allows to accurately model bubbly flows that involve two fluids with large density differences, for instance water and air, while only modelling the heavier fluid.

The structure of this paper goes as follows. 
Our methodology is presented in section \ref{methodology}, in which the overall algorithm of the particle finite element method is briefly presented first in section \ref{PFEMalgoSection}. 
The physical concepts of free surface flows are described in section \ref{ing1}, 
followed by the domain detection algorithm in section \ref{alphashapeSection}.
Next, section \ref{mesh} describes our approach to adapt and refine the mesh. 
We then focus on different boundary conditions in section \ref{bc}, one of which is the incompressibility condition of empty bubbles.
Our solver is finally verified on a few benchmark tests in section \ref{valid}, 
and a few additional results are shown in section \ref{results}.

\section{Methodology}\label{methodology}

The need for a method that seamlessly incorporates strong topological changes during a simulation 
is what led to the birth of the particle finite element method.
Whenever there exists a discontinuity in the parameters of a model (Young's modulus, latent heat, density, viscosity...), 
the solution of the problem or its derivatives may also present discontinuities that are called interfaces or fronts. 
These fronts are mobile, they can merge, split, nucleate or disappear. 

An accurate method that captures these moving fronts requires two main ingredients.
The first is a solver that numerically solves the physical equations of the model. 
The second is an approach to track the fronts, in our case the free surface.
These two ingredients are presented in this section, after briefly describing the overall PFEM algorithm.
\subsection{The Particle Finite Element Method}\label{PFEMalgoSection}

\begin{figure}
    \centering
    \includegraphics[width=.8 \textwidth]{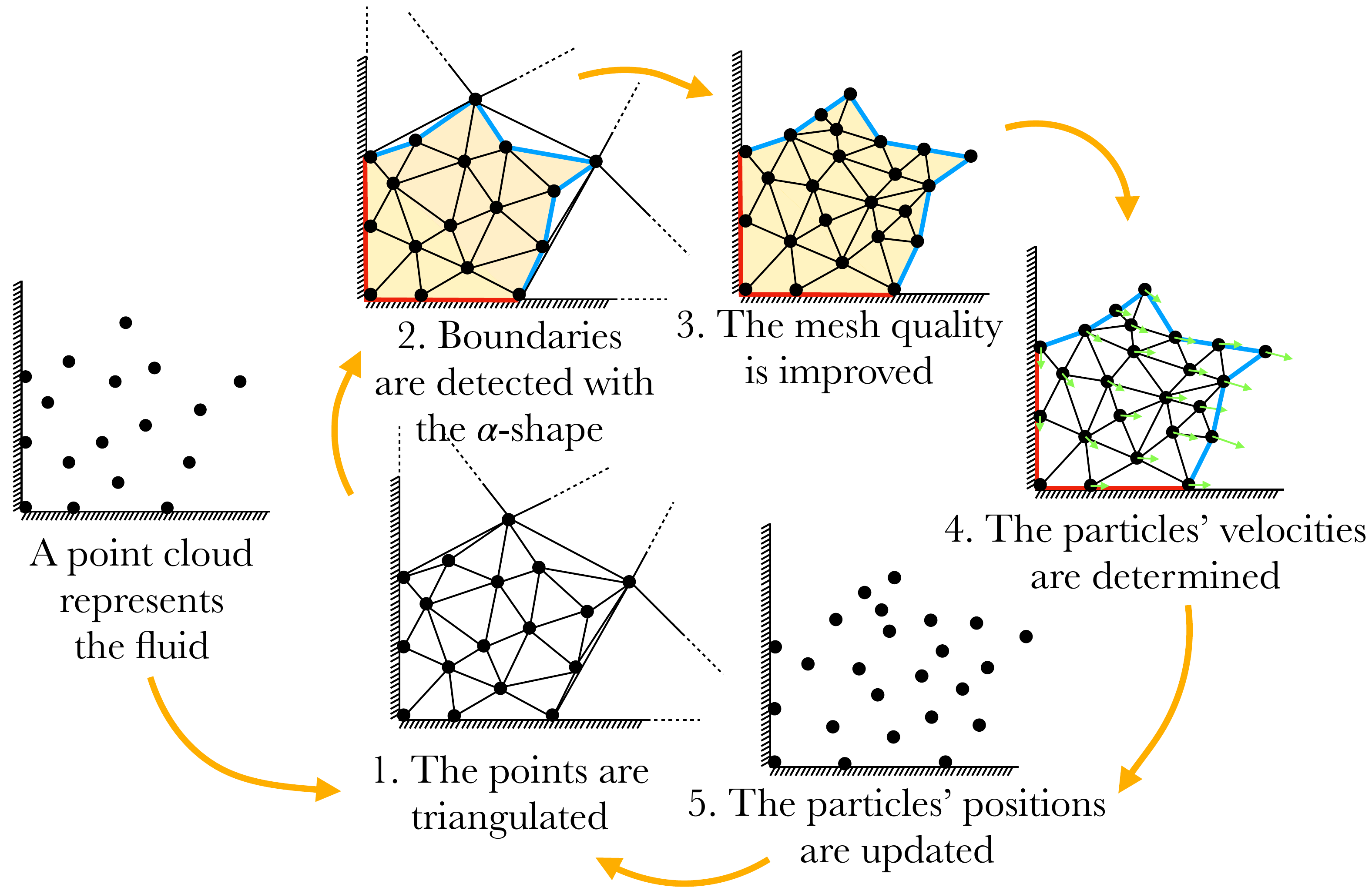}
    \caption{The main steps in the PFEM algorithm. 
    }\label{pfemAlgo}
\end{figure}

As its name suggests, the PFEM follows the evolution of a fluid by tracking the motion of particles.
Therefore, an initial particles distribution is generated. 
This can be obtained from a mesh that represents the initial state of the fluid.
From this mesh, the nodes are taken as the particles of the simulation.

Next, a geometry that defines the full computational domain is created. 
Geometrical points, defined as control nodes, are needed for the representation of the domain.
These control nodes are then connected to create the solid boundaries of the geometrical domain.

Having finished the preliminary steps, the main loop starts. 
The steps refer to those depicted in Figure \ref{pfemAlgo}.
First, the control nodes and fluid particles are triangulated together, and the solid boundary edges are constrained.
A complete triangulation of the computational domain is thereby obtained (step 1). 
Next, the $\alpha$-shape algorithm is applied to determine which elements belong to the fluid domain, 
also allowing to detect the solid and free surface boundaries (step 2). 
This step is detailed in section \ref{alphashapeSection}.
From the full geometrical domain, we determine which are the boundaries of the fluid along solid walls. 
Then, the quality of the fluid elements is improved using the Delaunay refinement algorithm (step 3). 
The algorithm also ensures that a size field is respected for all the elements in the fluid domain.
The mesh refinement algorithm is detailed in section \ref{mesh}.
The definition of the mesh is now complete: 
Elements are guaranteed to meet a certain quality and to respect the size field, and the boundaries are also clearly defined. 

Since some particles may have been added during the mesh adaptation process, it is necessary to project the velocity of the previous time step onto these particles. 
This is achieved using the deformed mesh of the previous time step using piecewise interpolations. 
The equations of motion can now be solved using the procedure described in section \ref{ing1} (step 4).
The material velocity is thereby obtained throughout the whole fluid domain, and in particular on the particles. 
Using this velocity, the particles are displaced (step 5). 
\subsection{Solving free surface flows}\label{ing1}
The PFEM solves the equations of motion using a finite element formulation. 
In this paper, we focus on incompressible free surface flows, hence the Navier-Stokes equations are used in their incompressible form. 
The strong form can be expressed as follows:
\begin{align}
    \rho \frac{D\mathbf u}{Dt} &= -\nabla p + \mu \nabla^2 \mathbf u + \rho \mathbf g\label{moment}\\
    \nabla \cdot \mathbf u &= 0.\label{mass}
\end{align}
Here, $\rho$ is the density of the fluid, $\mathbf u$ its velocity, $p$ the pressure, $\mu$ the dynamic viscosity,
and ${\mathbf g}$ the gravitational acceleration. 
The momentum equation \eqref{moment} is expressed with the material derivative of velocity. 
Indeed, recall that the PFEM is a Lagrangian method:
the velocity obtained at each node is the velocity of a material point within the fluid. 
The non-linearity of the Eulerian form of the momentum equation caused by the convective acceleration is not present in the PFEM.
This is a great advantage when solving the system of equations. 

\begin{figure}
    \centering
    \includegraphics[width = 0.6\textwidth]{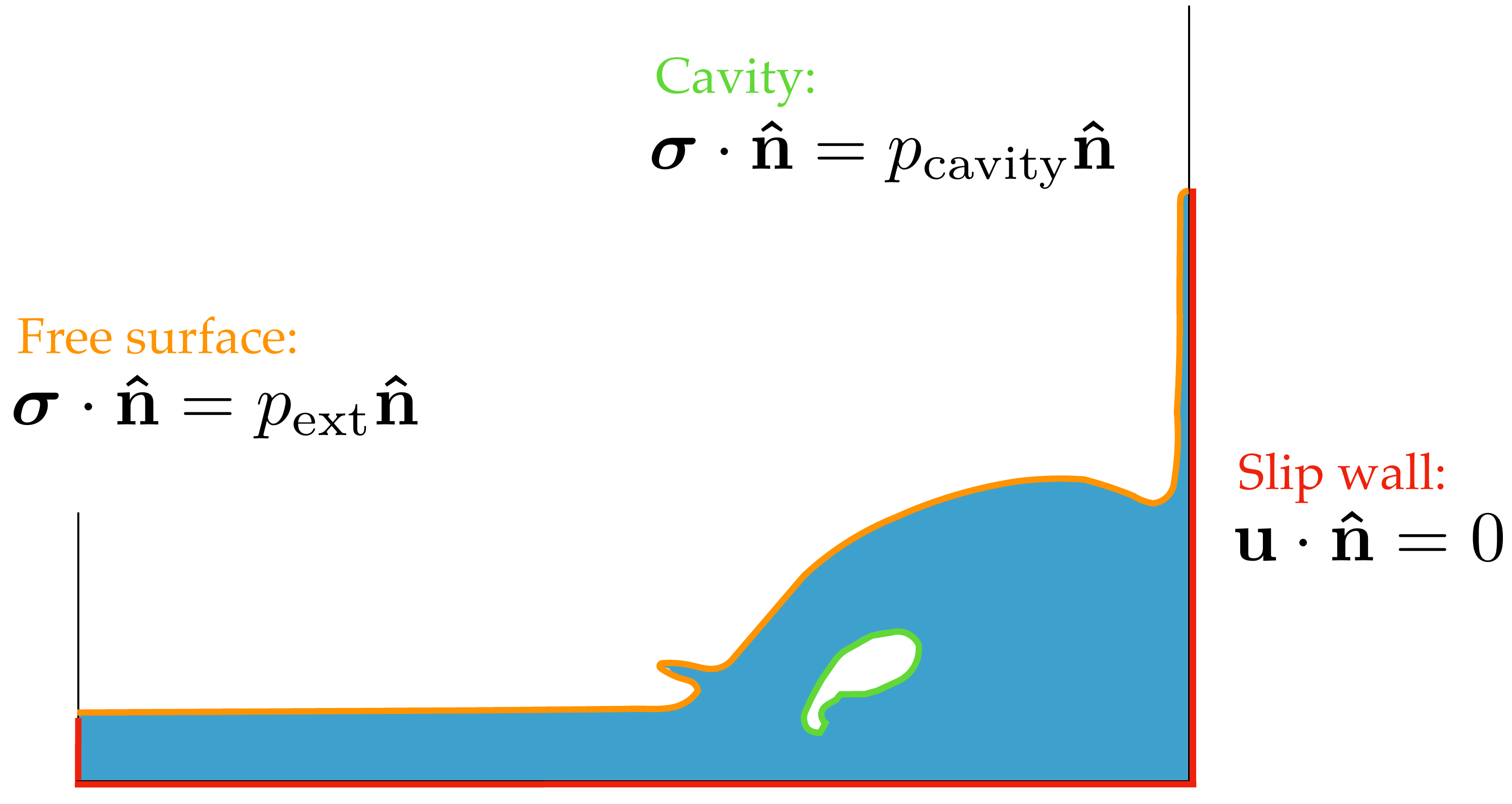}
    \caption{Boundary conditions along the fluid.}\label{bcFig}
\end{figure}

Equations \eqref{moment} and \eqref{mass} need to be complemented with boundary conditions. 
Along solid walls, a slip-wall condition is applied, meaning that the velocity's normal component to the wall is zero:
\begin{align}
    \mathbf u \cdot \mathbf {\hat n}\mathbf = 0 \text{\hspace{.5cm} for }\mathbf x \in \Gamma_{\text{wall}}.\label{slip}
\end{align}
For a detailed explanation of the slip-wall condition, we refer to section \ref{slipbc}.
Along the free surface, the stress normal to the boundary is subject to an exterior pressure: 
\begin{align}
    \boldsymbol{\sigma} \cdot \mathbf{\hat n} = p_{\text{ext}}\mathbf{\hat n} \text{\hspace{.5cm}} \mathbf x \in \Gamma_{\text{f.s.}},
\end{align}
where $\boldsymbol{\sigma}$ is the fluid stress tensor.
Finally, for cavities within the fluid, a pressure is applied around the bubbles to emulate the presence of an incompressible fluid within these cavities:
\begin{align}
    \boldsymbol{\sigma} \cdot \mathbf{\hat n} = p_{\text{cavity}}\mathbf{\hat n} \text{\hspace{.5cm}} \mathbf x \in \Gamma_{\text{bubble}}.
\end{align}
The derivation of this pressure $p_{\text{cavity}}$ is presented in section \ref{incompbc}.
An illustration of the boundary conditions is presented in Figure \ref{bcFig}. 

Equations \eqref{moment} and \eqref{mass} are solved with the fluid component of the Migflow solver\cite{migflow},
which implements the standard piecewise pressure stabilized linear finite element formulation\cite{pspg}.
Solving these equations, the velocity field is obtained throughout the domain.
In particular, the velocity vector at the nodes corresponds to the material velocity $\mathbf{U}_i^t$ of the particles.
The position $\mathbf{X}_i^t$ of the particles is then updated as follows:
$$
\mathbf{X}_i^{t+1} = \mathbf{X}_i^t + \mathbf{U}_i^t \Delta t, \text{\hspace{.5cm}} i= 1,...,n
$$
for each particle $i$.
\subsection{Using $\alpha$-shapes to capture the domain changes}\label{alphashapeSection}

\begin{figure}
    \centering
    \includegraphics[width = 0.99\textwidth]{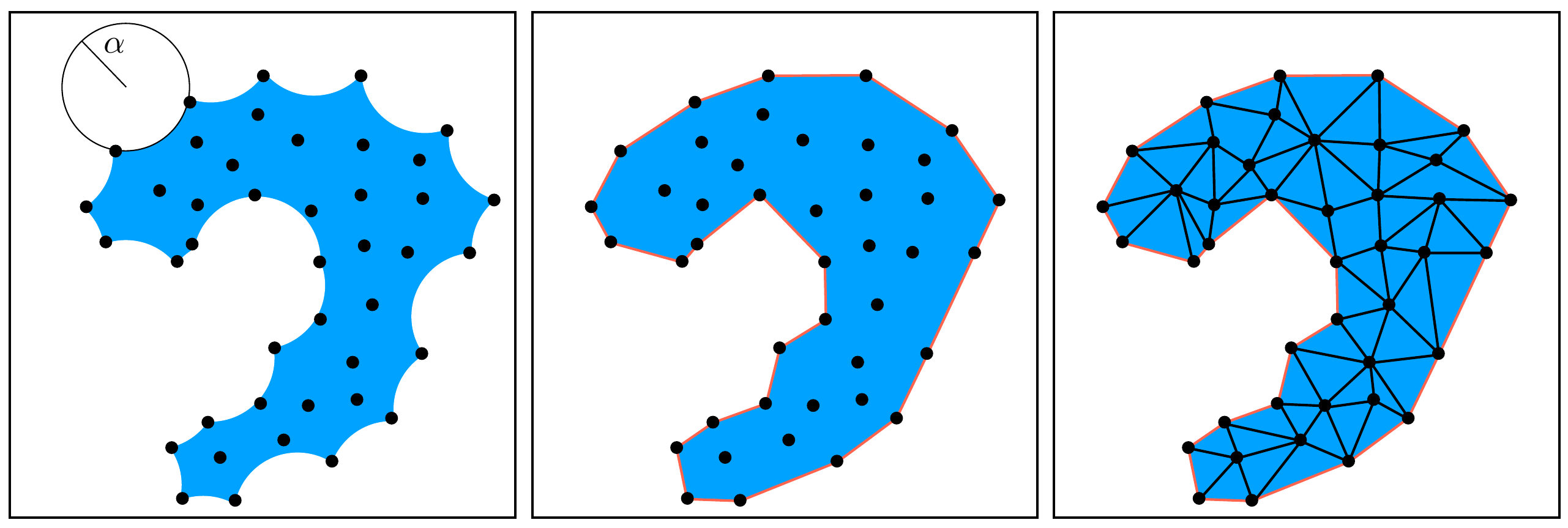}
    \caption{Left: $\alpha$-hull of a set of points. Center: the corresponding $\alpha$-shape. Right: The Delaunay triangulation of the $\alpha$-shape.}\label{alphahull}
\end{figure}

Due to the displacement of the particles, the domain is constantly undergoing geometrical and topological changes. 
The shape of the fluid domain therefore needs to be detected at every time step.
To this end, the approach proposed in the particle finite element method considers a shape detection algorithm closely related to the $\alpha$-shape of a set of points.  
Initially proposed by Edelsbrunner \cite{edelsbrunner}, the $\alpha$-shape is a generalization of the convex hull of a set of points $S$.  
Consider $\alpha$ a positive real number whose dimension is a length. 
An open disk of radius $\alpha$ is said to be empty if it contains no point of $S$. 
The $\alpha$-hull of S is the region that does not intersect any open disk of radius $\alpha$. 
The $\alpha$-shape has the same topology as the $\alpha$-hull but every circle arc 
of radius $\alpha$ of the $\alpha$-hull of $S$ is replaced by a straight line in the $\alpha$-shape of $S$ (see Figure \ref{alphahull}).

\begin{figure}
    \centering
    \includegraphics[width = 0.6\textwidth]{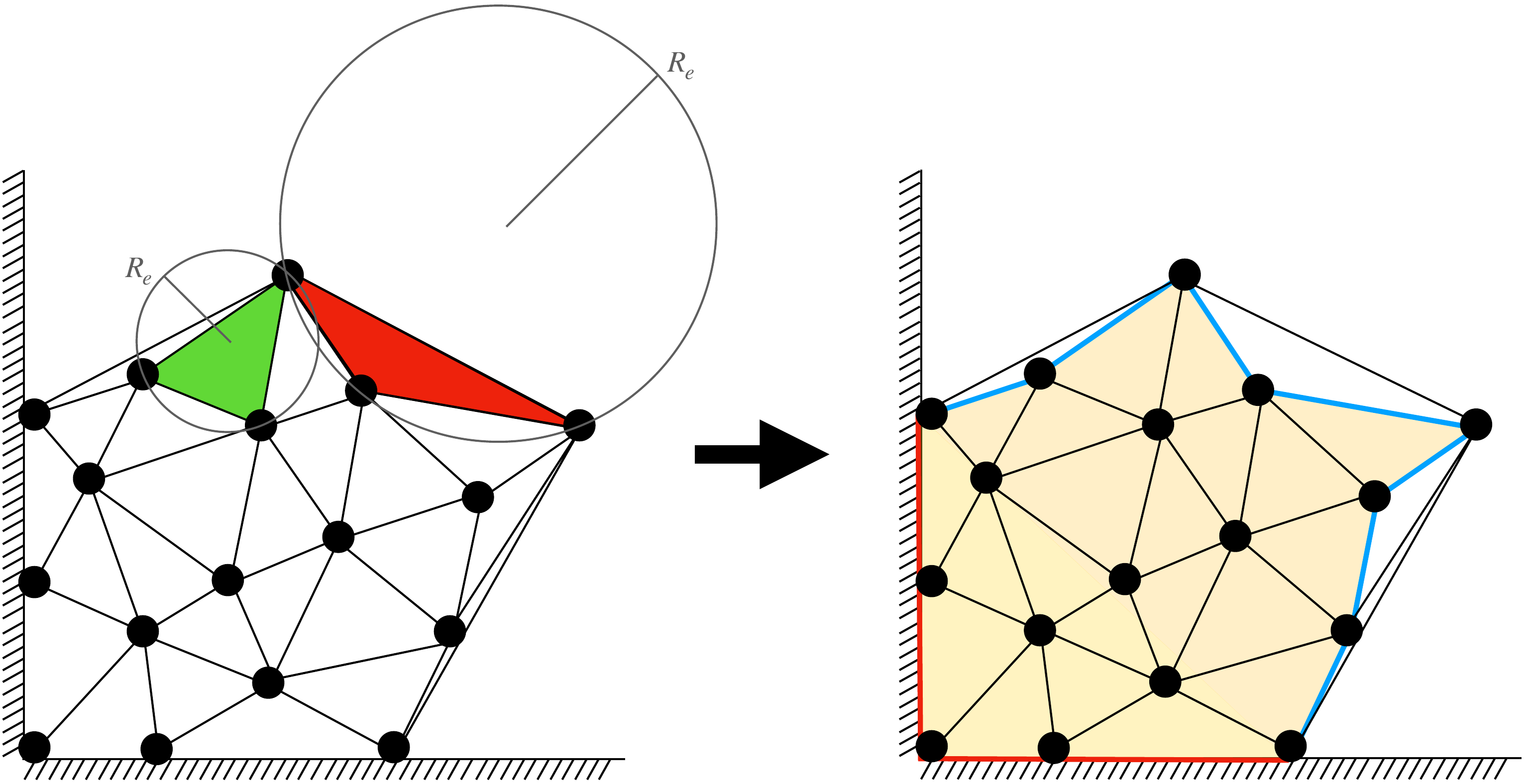}
    \caption{To build the $\alpha$-shape of the set of fluid particles, we build the Delaunay triangulation. 
    Only elements of the triangulation whose circumradii are smaller than a prescribed value are considered part of the fluid.
    The green element is added to the fluid, while the red element is not.
    Boundaries are naturally detected by the algorithm.}\label{alphashape}
\end{figure}

The Delaunay triangulation $\mathcal{T}(S)$ of $S$ is the easiest way to build the $\alpha$-shape of $S$.
We remove from $\mathcal{T}(S)$ triangles that have a circumcircle greater than $\alpha$ and the boundary of that 
domain is the $\alpha$-shape of $S$. 
In this work, we assume that $\alpha$ is a field. 
In other words, $\alpha(\mathbf x)$ depends on the position $\mathbf x$.  
Assuming that $\mathbf x_e$ is the circumcenter of element $e$, we define an adimensional parameter
$$ \alpha_e = \frac{R_e}{\alpha(\mathbf x_e)},$$
where $R_e$ is the circumradius of element $e$. 
Triangles having a value of $\alpha_e$ greater than a certain threshold value (typically, a value of 1.2) are removed from the fluid domain (see Figure \ref{alphashape}). 
The definition of $\alpha(\mathbf x)$ is detailed in section \ref{mesh}. 

\begin{figure}
    \centering
    \includegraphics[width=.8\textwidth]{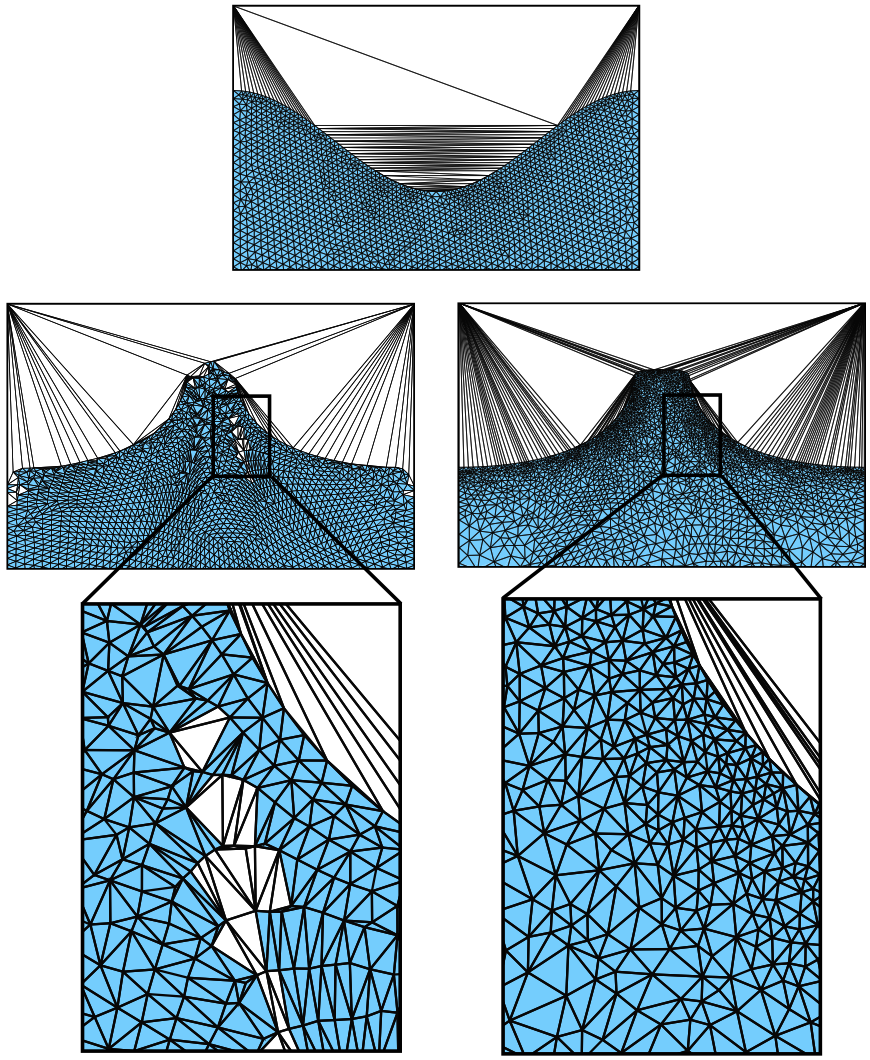}
    \caption{Comparison of a sloshing simulation with (right) and without (left) mesh adaptation. 
             Holes appear inside the mesh due the $\alpha$-shape algorithm that removes elements that no longer respect the quality and size constraints.}\label{alphashape2}
\end{figure}

We combine the $\alpha$-shape approach with a finite element method to solve fluid flows. 
The initial set of points $S_0$ at $t=0$ corresponds to the user-defined finite element mesh of a given fluid domain $\Omega$. 
The elements of this initial mesh must have circumcircles that respect the size field $\alpha(\mathbf x)$.
To generate this initial mesh, we use a \emph{Delaunay-based} mesh generator (Gmsh \cite{gmsh}). 
In a classical mesh generation process, points are inserted as part of a \emph{Delaunay refinement} process with the aim of robustly respecting the mesh size, such that the circumradius of the triangles is controlled. 
The Delaunay triangulation does not necessarily minimize the length of the edges. 
In other words, the Delaunay triangulation does not always connect points that are close to each other. 
The only result in that regard is that,
in a Delaunay triangulation, the graph distance between two points is never more that 2 times the Euclidian distance between those points. Thus, if points are inserted randomly, 
even in a controlled process such as a Poisson process, there is no guarantee that large triangles will not appear in $\mathcal{T}(S)$. 
Slightly disturbing the set of points by a distance locally smaller than $\alpha(\mathbf x)$ is not really a concern: 
the Voronoi diagram (dual of the Delaunay triangulation) is a very stable structure and the topology of the $\alpha$-shape will not be disturbed unless the topology of the fluid domain changes \cite{edelsbrunner}. 
However, if nothing is done for a large number of time steps and the points are simply moved using the fluid velocity, then the initial structure proposed by the mesh generator will be broken and major disturbances in the $\alpha$-shape will appear: 
holes in the fluid domain, oscillatory and chainsaw-like structures along the free surface.
In Figure \ref{alphashape2}, elements in blue are those recognized by the $\alpha$-shape as fluid. 
As the simulation advances, the initial distribution of the particles is disturbed by their velocity.
Without any mesh adaptation, as presented in the left column, holes appear within the fluid domain.
Elements which should have remained part of the fluid have been removed, 
since they no longer respect the $\alpha$ criterion described in the previous section.
In the PFEM, it is therefore crucial to adapt the mesh through particles insertion and removal.

The philosophy behind our approach is as follows. 
We assume that at time step $t$ the set of points $S_t$ has a structure resulting from a clean Delaunay refinement. 
We therefore determine the fluid domain's shape through the $\alpha$-shape process described above. 
Next, in order to ensure that all elements within the fluid domain have sufficient quality, the set of points $S_{t}$ is modified by a Delaunay refinement algorithm with a structure of proof to restore a clean point distribution.
Then, we calculate the velocities ${\bf U}_i^t$ of the points by finite elements on the domain bounded by the $\alpha$-shape of $S_t$.  
We then move the points according to the fluid velocity to create $S_{t+\Delta t}$. 
The time step $\Delta t$ must not exceed a Courant–Friedrichs–Lewy (CFL) condition of 1: $$\Delta t < \inf_{\mathbf x} \frac{\alpha(\mathbf x)}{\|{\bf{U}^t(x)\|}}.$$
This is perfectly acceptable since we assume that the mesh size chosen corresponds to the characteristic length of the given phenomena. 
\subsection{Mesh adaptation}\label{mesh}

As shown in the right column of Figure \ref{alphashape2}, in contrast to the left column, quality-based mesh adaptation allows to accurately represent the evolution of the fluid domain.
In this section, we describe the mesh adaptation algorithm which, along with the modified version of the $\alpha$-shape algorithm described above, ensures an accurate representation of the domain at each time step. 
An example using a spatially variable analytic incompressible velocity field is then used to demonstrate the effectiveness of the proposed approach.

\subsubsection{Adaptive meshing algorithm}\label{meshalgo}

\begin{figure}
    \centering
    \includegraphics[width=.6\textwidth]{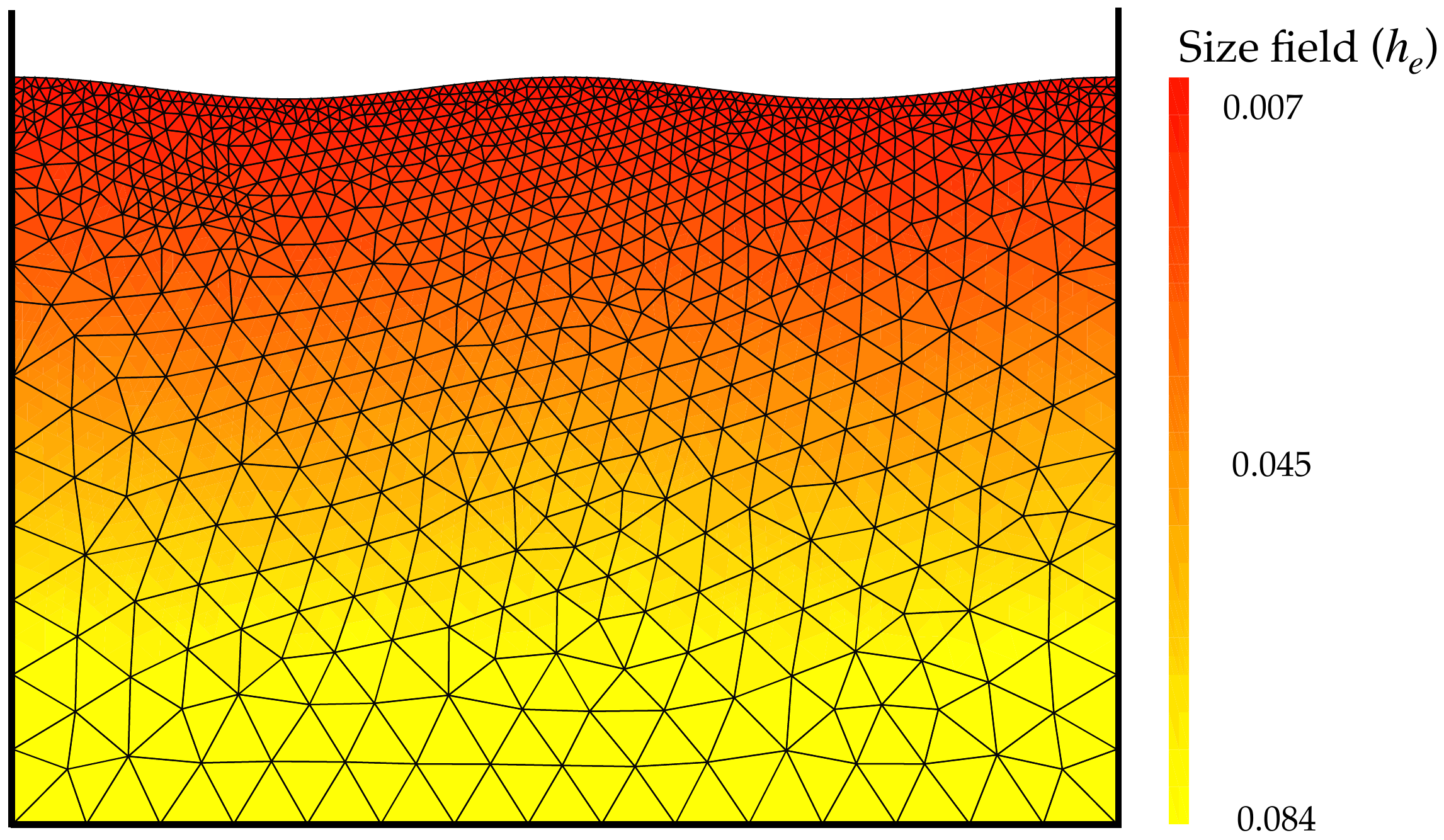}
    \caption{The size of the elements is dictated by their distance to the free surface.}\label{sizeField}
\end{figure}

First, a size field is defined to allow variations in the size of the mesh. 
For instance, a higher resolution may be desired near the free surface to accurately simulate the strong deformations and topological changes. 
A KD-tree structure is used to compute the distance to the free surface to derive this size field. 
The position of the free surface at time $t+\Delta t$ is determined by advecting the free surface at time $t$ with the velocity field.
An example is presented in Figure \ref{sizeField}.

To adapt the mesh based on this size field, nodes are inserted into the triangulation.
To this end, an algorithm greatly inspired by Chew’s algorithm for Delaunay mesh refinement is employed \cite{chew}. 
Chew’s algorithm ensures that all triangles within the triangulation are well-sized and well-shaped and that internal and external boundaries are maintained. 
Shewchuck has since demonstrated that using this algorithm, no angles smaller than $26.5^o$ will be generated, 
confirming the well-shapedness of the mesh resulting from this algorithm\cite{shewchuk}.
The user-defined size field dictates the size of the elements.

Note that the algorithm is performed only after the fluid domain has been detected.
Only elements which have been defined as being part of the fluid by the $\alpha$-shape will potentially be adapted.
Hence, only the part of the mesh defined as the fluid domain will be improved in terms of quality from one time step to another. 

The algorithm takes a constrained Delaunay triangulation $\mathcal{T}_c(S)$ as input. 
$\mathcal{T}_c(S)$ is a Delaunay triangulation in which certain edges are constrained, 
meaning that they must remain in the mesh and therefore cannot be flipped.
In practice, the constrained edges are those belonging to the solid boundaries of the domain 
as well as the edges that have been detected as free surface edges by the $\alpha$-shape.
With $\mathcal{T}_c(S)$ as input, the algorithm then seeks to meet the imposed quality and size criteria.

\begin{figure}
    \centering
    \includegraphics[width = .9 \textwidth]{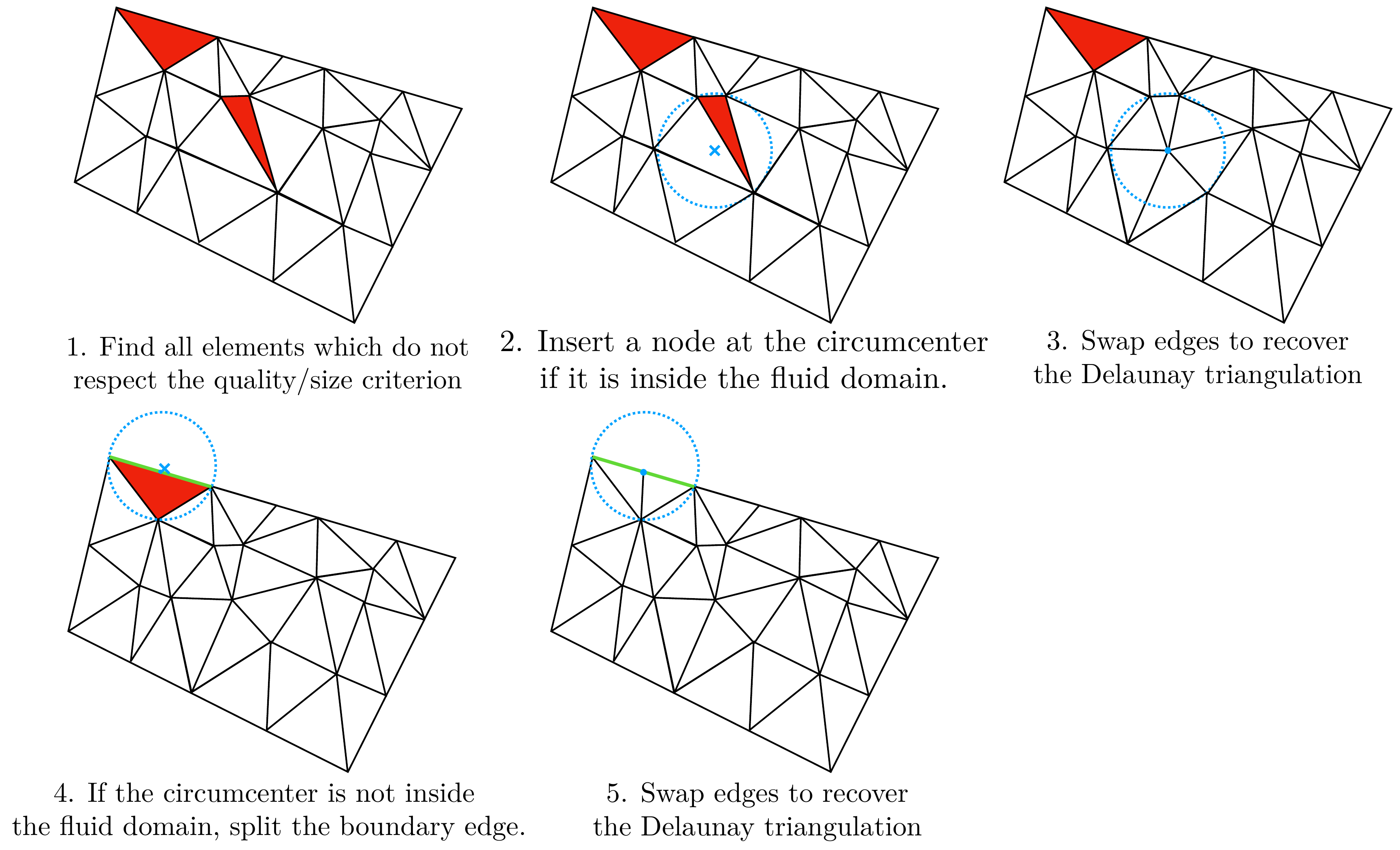}
    \caption{Main steps of the mesh adaptation algorithm.}\label{algoIllustration}
\end{figure}

The essence of the approach consists of inserting nodes in the triangulation at the circumcenter of poorly shaped or poorly sized triangles. 
Edges are flipped after insertion to ensure the mesh remains Delaunay.
In case a circumcenter falls outside the meshing domain, then the bounding edge that crosses the line-of-sight of the triangle to its circumcenter is split. 
Algorithm \ref{algo} summarizes the process. 
Figure \ref{algoIllustration} illustrates the main steps of the algorithm on a small sample of elements.

Node insertion at the circumcenter of poor-quality elements guarantees overall mesh improvement. 
Indeed, the only new edges created by an insertion of a node are edges connected to this node (see for example Figure \ref{algoIllustration}, step 2 to 3).
This node has been inserted at the circumcenter of a triangle which respects the Delaunay criterion.
In other words, there are no other nodes within the circumcircle of this triangle.
Therefore, none of the new edges can be shorter than the circumradius of this triangle.
This avoids the appearance of triangles with a small innner radius to circumradius ratio.
This property of the algorithm is essential to the PFEM mesh adaptation step. 

\begin{algorithm}
    \caption{Mesh adaptation}\label{algo}
    \begin{algorithmic}
        \State Get the $\alpha$-shape of the point cloud and define the fluid domain;
        \State Build $\mathcal{T}_c(S)$ of the fluid domain;
        \State Collapse internal edges that are too small with respect to the size field;
        \State Get all the elements in the $\alpha$-shape that do not meet the quality or size criterion;
        \While {There are still poorly shaped or sized elements}
            \State Choose the largest element that does not respect the quality criterion;
            \State Determine its circumcenter;
            \If {The circumcenter is inside the fluid domain}
                \State Insert a new node at the circumcenter;
            \EndIf
            \If {The circumcenter is outside of the fluid domain}
                \State Insert a node at the middle of the constrained edge that blocks the triangle from seeing its circumcenter;
                \State Delete any node that is closer to the newly created node than the length of the new edge;
                \EndIf
            \State Do edge swaps to recover a Delaunay mesh;
        \EndWhile
    \end{algorithmic}
\end{algorithm}

The quality criterion for an element $e$ is defined as $ \gamma_e = r_e/R_e$, 
where $r_e$ is the inscribed radius and $R_e$ is the circumscribed radius.
The size criterion compares the circumradius of the element to the prescribed size field. 
Before the mesh refinement algorithm runs, a coarsening step is performed. 
This allows to avoid excessively small elements to remain inside the mesh, where the size field actually allows a coarser mesh. 
If two particles are too close to each other, their mutual edge is collapsed and $\mathcal{T}_c(S)$ is updated accordingly.
Finally, to ensure termination of the algorithm, no points are inserted for elements whose circumradius is smaller than a limit value.

Using the mesh adaptation algorithm in the PFEM greatly improves the quality of the mesh. 
This step turns out to be crucial, as it removes uncertainties linked to the $\alpha$-shape’s decision of the shape of the fluid. 

\subsubsection{Adaptive meshing illustration: vortex-in-a-box}
To illustrate the ability of the mesh adaptation algorithm to capture flow features down to the size of the mesh, the vortex-in-a-box simulation is presented below. 
In this flow, a disk is inserted into a circular flow inside a box with the following stream function: 
$$ \psi = \frac{1}{\pi} \sin^2(\pi x)\sin^2 (\pi y) \sin(2 \pi t / T)$$
The resulting velocity field is obtained as follows : 
\begin{align*}
    &u_x = \frac{\partial \psi}{\partial y} = \sin(2 \pi y) \sin^2(\pi x) \sin(2 \pi t / T)\\
    &u_x = -\frac{\partial \psi}{\partial x} = - \sin(2 \pi x) \sin^2(\pi y)\sin(2 \pi t / T)
\end{align*}
The domain is a box of size $[0,1]\times[0,1]$, and the initial disk is located at $(0.5, 0.75)$.
Gradually, as the disk follows the rotation of the flow, its shape evolves into a thin filament.
The filament can become infinitely long and thin, and the challenge of the numerical solver is to capture this filament down to the size of the grid. 
As a means of validation, a time-dependent term, $\sin(2\pi t/ T)$, is added to the velocity field that reverses the direction of the flow after a half period $T/2$. 
Hence, if the initial disk is recovered after a full period $T$, this shows that there is no area loss. 
Two different test cases are presented.
\begin{figure}
    \centering
    \includegraphics[width = .8 \textwidth]{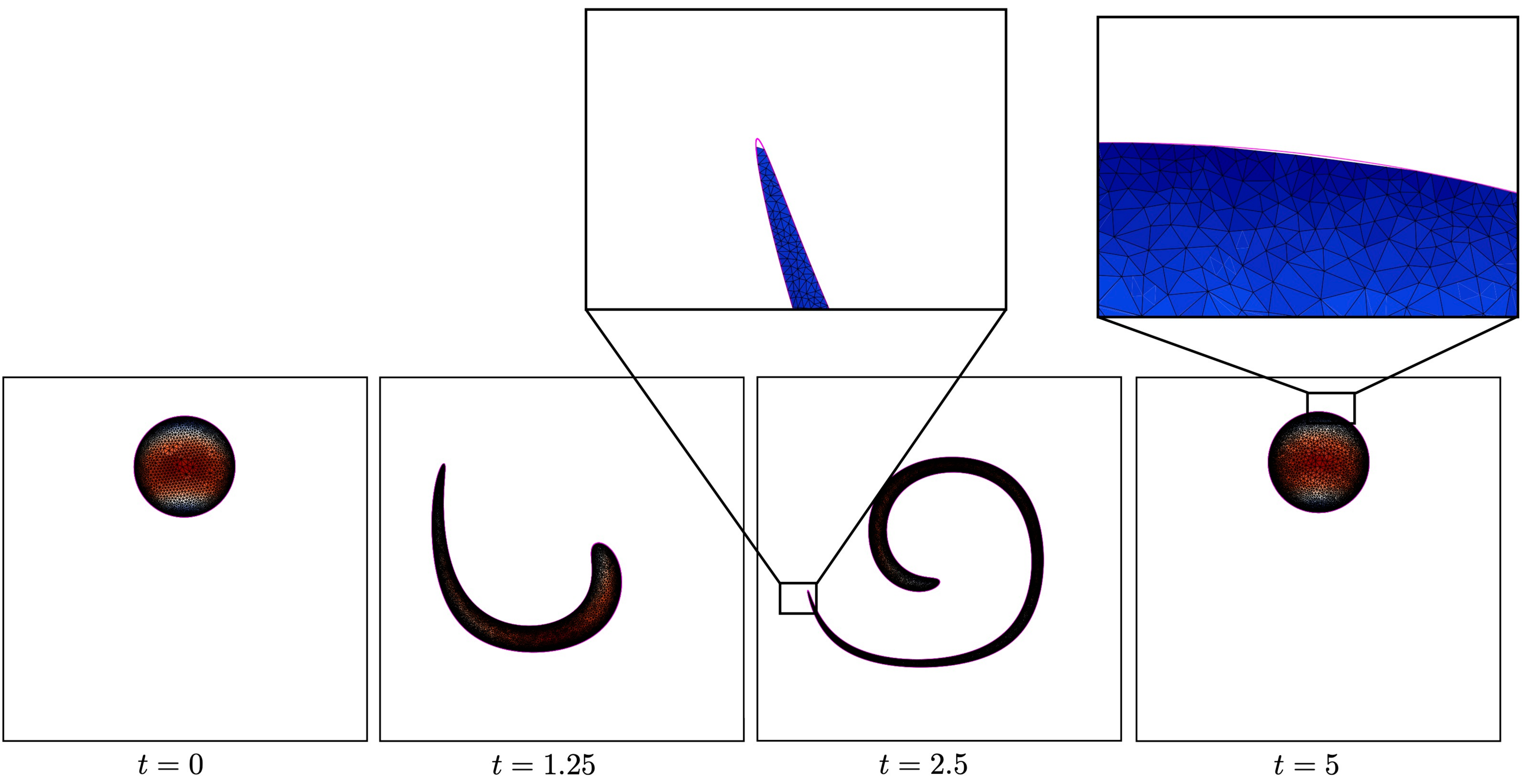}
    \caption{Snapshots of the vortex-in-a-box simulation for the first case (a bubble of fluid).}\label{vortex1}
\end{figure}
\begin{figure}
    \centering
    \includegraphics[width = .8 \textwidth]{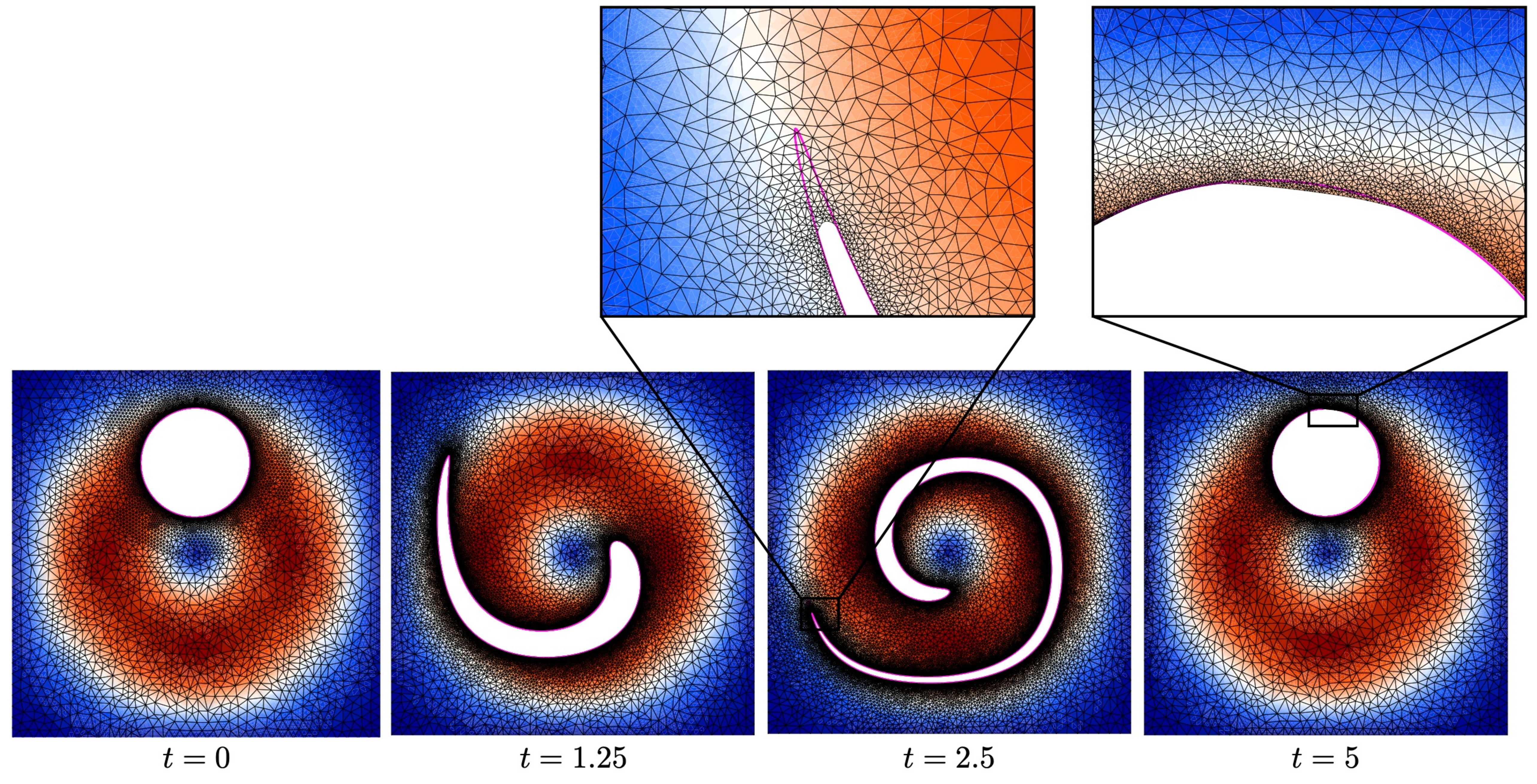}
    \caption{Snapshots of the vortex-in-a-box simulation for the second case (a hole in the fluid).}\label{vortex2}
\end{figure}

In the first, the fluid domain is considered as the inside of the disk (i.e., a bubble of fluid).
The second, its counter-part, considers the region around the disk as fluid (i.e., a hole in the fluid).
Time integration, for this particular example, is performed using a fourth order Runge Kutta scheme. 
Figures \ref{vortex1} and \ref{vortex2} present a few timeshots of the two different simulations.
The pink lines correspond to the actual position of the interface.

\begin{figure}
    \centering
    \includegraphics[width = .8 \textwidth]{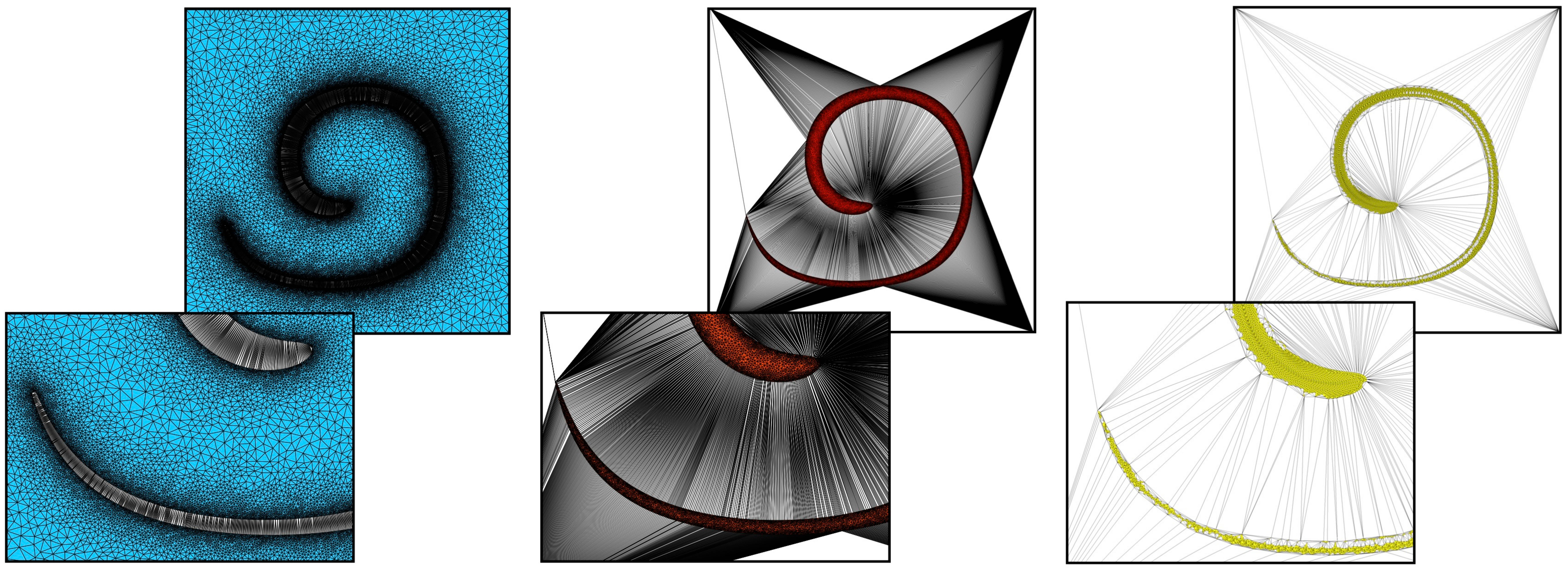}
    \caption{Difference between simulating the outside (left) or inside (middle) of the disk. 
            Comparison with a simulation without any mesh adaptation (right).}\label{vortex3}
\end{figure}

\begin{figure}
    \centering
    \includegraphics[width = .98 \textwidth]{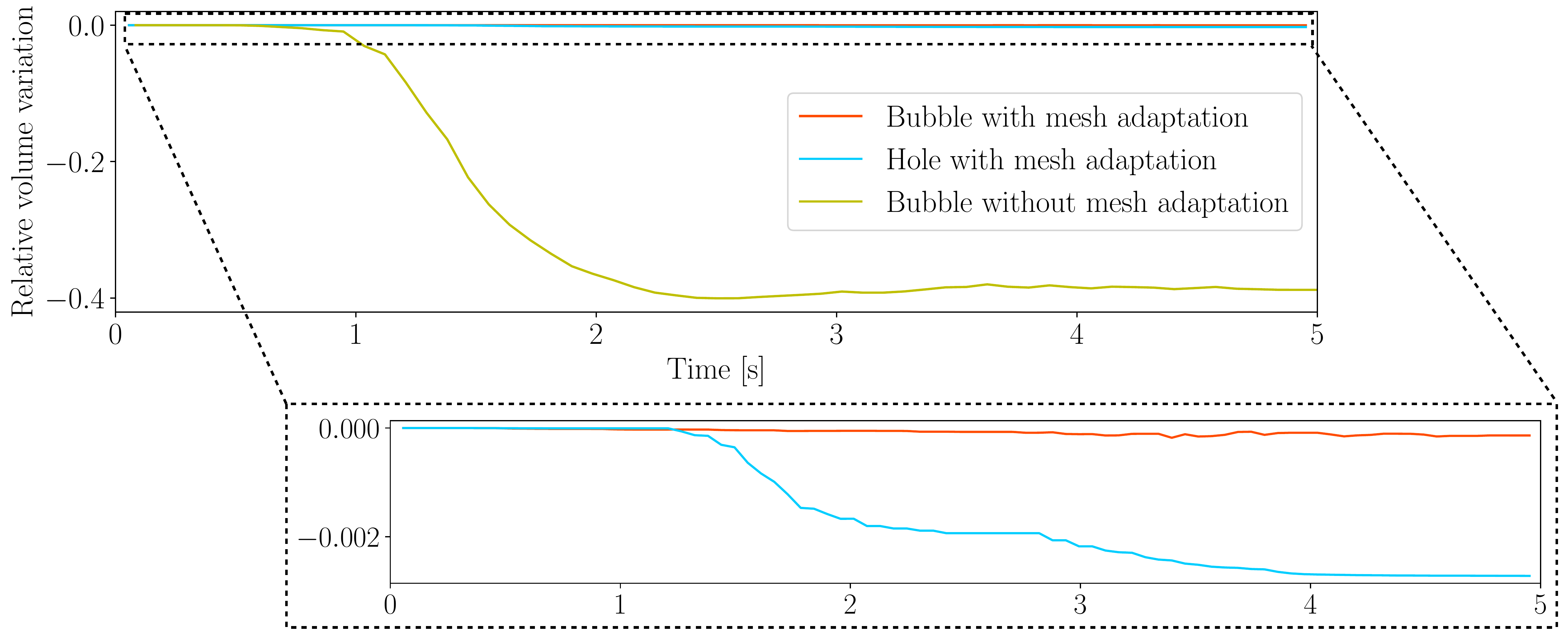}
    \caption{Volume variation during one full period of duration $T=5$s, when simulating the inside of the disk (bubble) and the outside of the disk (hole), 
            compared to a simulation without any mesh adaptation.
            For the sake of comparison, the relative volume variation of the \emph{inside} of the disk is computed for all three cases.}\label{vortex4}
\end{figure}

Figure \ref{vortex3} presents the full meshed domain, for both cases. 
Clearly, mesh adaptation is only performed inside the region that is considered to be the fluid region. 
A comparison is also done with a simulation without any mesh adaptation.
The number of elements is about the same ($\approx$ 25000) , but a uniform mesh size is considered throughout the domain. 
Figure \ref{vortex4} shows the volume variation of the disk over time, for all three cases presented above. 
For an objective comparison, the volume of the inside of the bubble is computed. 
The greater volume variation for the hole case is explained by the fact that the end of the filament becomes infinitely thin as the vortex rotates. 
Hence, capturing this thin filament with the $\alpha$-shape would also require infinitely small elements in this region. 
A portion of the volume of the disk is therefore lost as time is reversed.
The third image (at $t=2.5$s) of Figure \ref{vortex2} illustrates this phenomenon.
This situation does not happen when simulating the inside of the disk since there are no non-fluid elements of small, $\alpha$-acceptable size in this region. 
Only the small elements which become too distorted end up being removed by the $\alpha$-shape, which explains the very small variation of volume in this case.
Overall, it is clear that mesh adaptation is crucial to correctly track the evolution of the fluid domain.


\subsection{Boundary conditions}\label{bc}
Two boundary condition formulations are presented next. 
The first concerns the slip boundary condition along solid walls. 
The second adresses empty cavities within the fluid. 

\subsubsection{Slip boundary conditions}\label{slipbc}
Accurately capturing the boundary layer along a wall requires strong refinement along these walls, resulting in expensive computational efforts.
To circumvent this, it is common to allow a relative motion between the fluid and the solid wall, resulting in a slip-wall boundary condition.
Different approaches for slip boundaries in the PFEM have been proposed. 
Cerquaglia et al. \cite{cerquaglia} use one layer of elements along the solid wall to represent a boundary layer.
Others, for instance Cremonesi et al. \cite{cremonesiEulerBC} use an Eulerian-Lagrangian formulation to account for slip walls.

In this work, we suggest to avoid the use of fixed particles attached to the walls.
The boundaries of the computational domain are only represented geometrically.
To implement the slip boundary condition, the motion of particles situated along a solid wall is restricted to the direction tangential to this wall. 
In other words, the fluid particles slide along the wall and cannot reenter the bulk of the domain.
Any slip boundary condition, ranging from no-slip to free-slip, can then be applied with wall models.

\begin{figure}
    \centering
    \includegraphics[width=.5\textwidth]{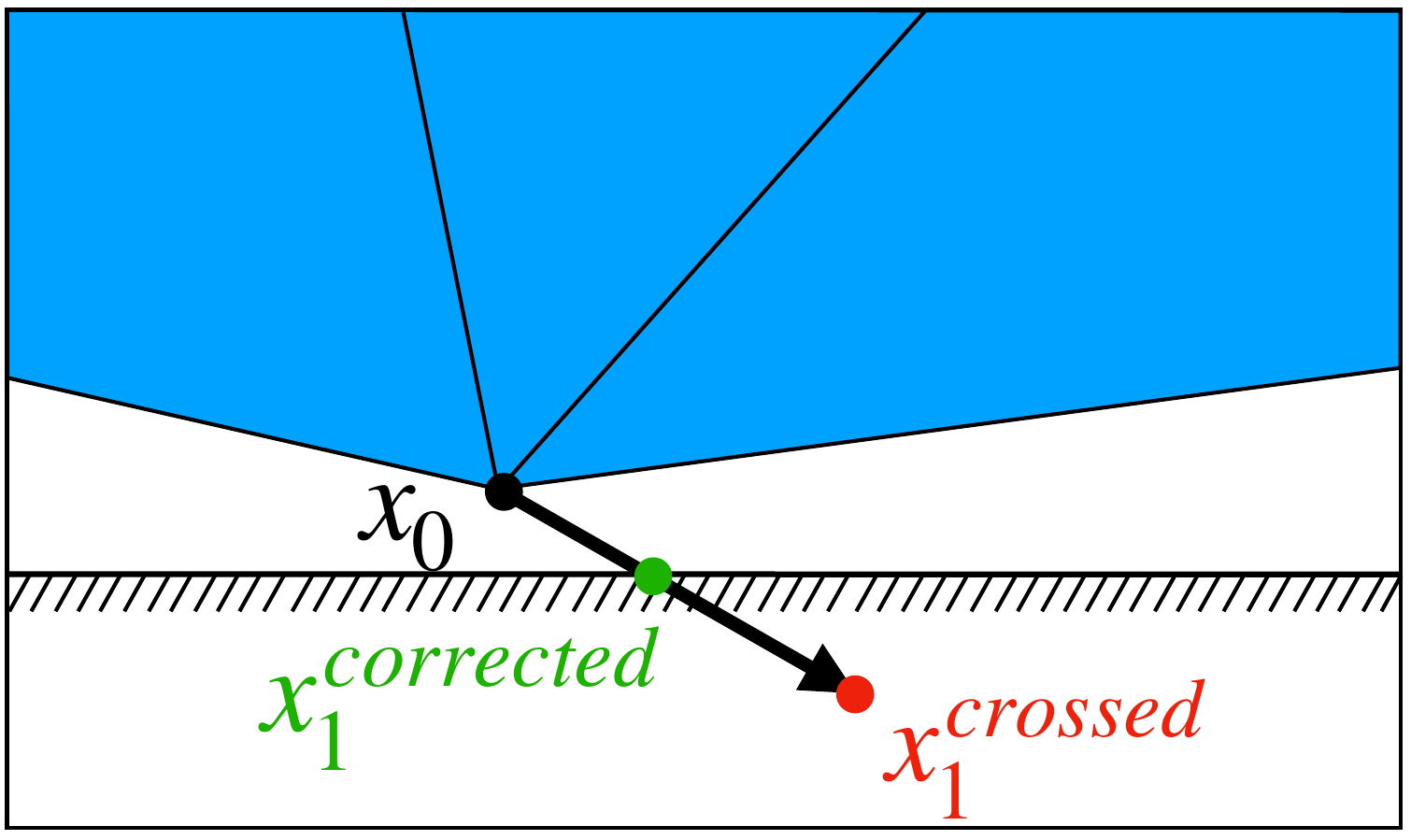}
    \caption{When a fluid particle crosses a solid wall, it is reprojected back onto the wall, and its velocity component normal to the wall is set to zero.}\label{reproj}
\end{figure}

A special case must be considered for particles that are still in the bulk, and cross a wall after their explicit position update, as illustrated in Figure \ref{reproj}.
Indeed, in such cases, the particle does not know \textit{a priori} that it will hit a wall.
To correct this, these particles are projected again onto the wall by backtracing their trajectory, and their velocity component normal to the wall is set to zero. 
The slip-boundary condition is thereby recovered. 
An error on mass conservation of the order of the time step is conceded in these situations since the incompressibility condition is locally disregarded.
A future improvement involves using an implicit update on the position in order to predict collisions between the fluid and solid walls.

\subsubsection{Empty bubble boundary condition}\label{incompbc}
\begin{figure}
    \centering
    \includegraphics[width=.8\textwidth]{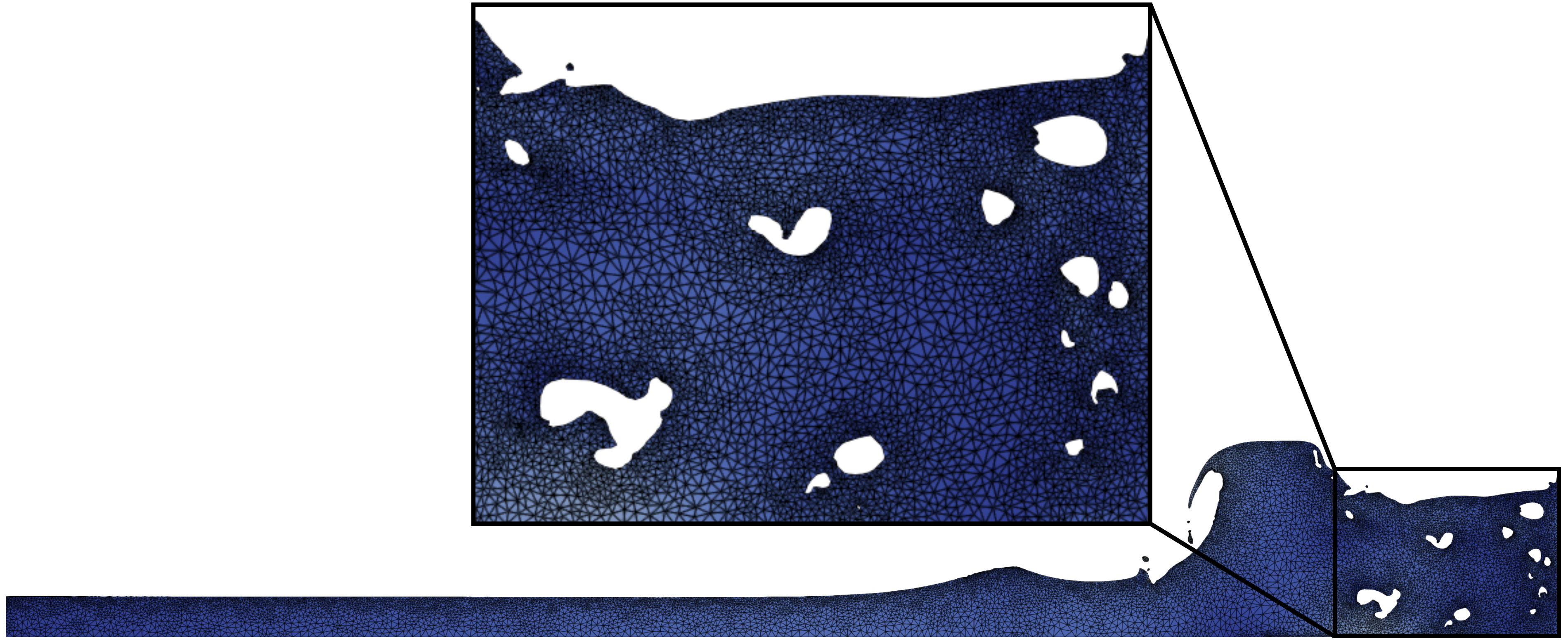}
    \caption{This highly transient flow generates the appearance of bubbles of air within the fluid.}\label{wetbedZoom}
\end{figure}

In some simulations, for example those driven by high levels of inertia, the fluid can encapsulate cavities to form bubbles, as illustrated in Figure \ref{wetbedZoom}.
Although the presence of these bubbles is natural, if no special attention is paid to them, they disappear.
If we ignore the fluid inside the bubbles, the model considers those regions as a void spaces, and fills them up.
Global mass conservation is imposed on the fluid, but not on the bubbles. 

Let us assume that the dynamics inside each bubble occur on a time scale that is much faster than that of the fluid. 
In other words, a variation of the fluid is instantaneously compensated inside the bubble.  
This is the case for a sufficiently high density ratio between the fluid and the bubble.
Between water and air, for instance, this assumption is reasonable. 
In this case, without modelling explicitly its flow, we can still consider the presence of another incompressible fluid inside the bubble.

We apply a boundary condition that exerts a pressure enforcing the incompressibility of the bubble, while taking into account its weight and the surface stresses on the interface.
The pressure term applied on this boundary is composed of three terms: a vertical hydrostatic pressure gradient that accounts for buoyancy, surface tension, and a constant term that enforces incompressibility:
\begin{align}
    p_{\text{cavity}}(\mathbf x) =  \underbrace{\rho_{\text{c}} g z}_{\text{buyoancy}} +  \underbrace{\kappa(\mathbf x) \sigma}_{\text{surface tension}} \label{p} + \underbrace{p_{\text{i}}}_{\text{incompressibility}}.
\end{align}
Here, $\rho_\text{c}$ is the density of the fluid inside the bubble, $g$ the gravitational acceleration, $z$ the vertical component of $\mathbf x$, $\kappa$ is the curvature, and $\sigma$ the surface tension coefficient.
These first two terms control the gradient of $p_{\text{cavity}}$ and can be explicitly computed.  

To control the average pressure $p_\text{i}$, we add an incompressibility constraint for each bubble:
\begin{align*}
    \int_{\Omega_{c}} \nabla \cdot \mathbf{u}_c ~dV &= 0\nonumber \Longleftrightarrow \int_{\Gamma_{c}} \mathbf{u}_c \cdot \hat{\mathbf n}~ dS = 0, 
\end{align*}
where $\mathbf{u}_c$ is the velocity of the fluid inside the bubble.
By continuity, the velocity of both fluids is the same at the interface. 
Hence, we can express this condition only with the velocity of the modelled fluid, 
\begin{align}
    \int_{\Gamma_c} \mathbf u \cdot \mathbf{\hat n} ~dS = 0.\label{incompp}
\end{align}
For each bubble we have thus inserted one new equation, the incompressibility constraint \eqref{incompp}, and one new degree of freedom $p_\text{i}$.
Those can be seen as Lagrange multipliers used to enforce the constraints linking the nodes on the boundary of each bubble.
A validation test case of this boundary condition is presented in section \ref{validBubble}.
\section{Verifications}\label{valid}
This section aims at verifying the different features of our PFEM solver.
Two test cases are considered.
First, we consider a low amplitude fluid sloshing in a tank. 
This configuration is endowed with an analytical solution so we can verify analytically the overall approach. 
The second example is a rising bubble simulation. Other authors have used that test case so we can compare our
results with references and verify the accuracy of our incompressibility condition around cavities. 

\subsection{Low amplitude sloshing}
\begin{figure}
    \centering
    \includegraphics[width=.95\textwidth]{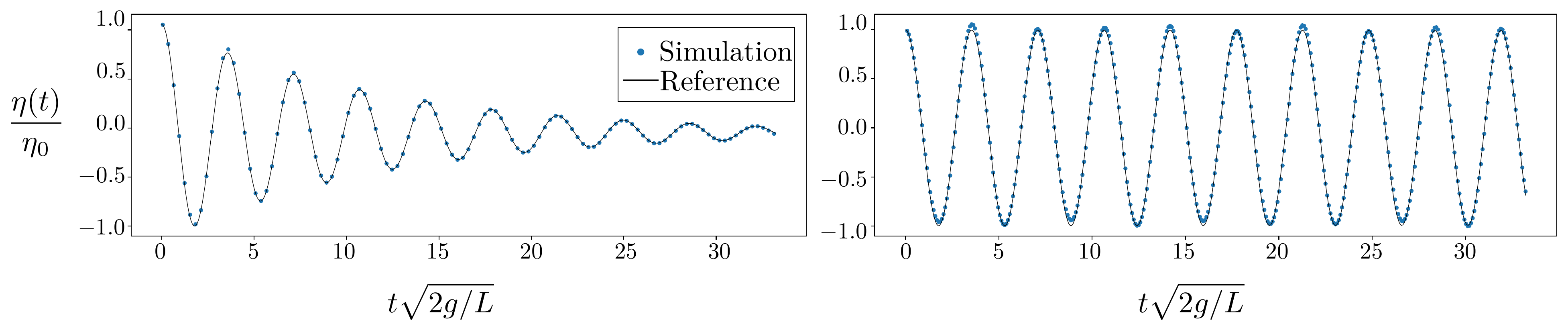}
    \caption{Evolution of wave height over time in a low amplitude sloshing experiment. 
    Left: dampening due to viscous effects. Right: no viscous effects.}\label{sloshingGraph}
\end{figure}

The sloshing experiment consists in analysing the evolution of a fluid in a container, 
with an initial perturbation of the height of the fluid. 
This initial height takes the form of a sinusoidal wave:
\begin{align*}
    h(x,0) = h_0 + \eta(x,0) = h_0 + \eta_0 \cos (2\pi x/L),
\end{align*}
where $h_0$ is the mean height of the fluid in the container, $\eta_0$ is the amplitude of the perturbation, 
and $L$ is the length of the container.
For sufficiently low values of $\eta_0$, we can assume a linearisation of the Navier-Stokes equations to obtain
an analytical solution. 
Indeed, the evolution of the perturbation is given by \cite{wuSloshing}:
\begin{align*}
    \frac{\eta(t)}{\eta_0} = 1 - \frac{1}{1+4\nu^2k^3/g}\Big[ 1 - e^{-2\nu k^2 t}\Big( \cos\sqrt{kg}t + 2 \nu k^2 \frac{\sin \sqrt{kg}t}{\sqrt{kg}}\Big) \Big],
\end{align*}
where $\nu$ is the kinematic viscosity of the fluid and $k=\pi/L$ is the wave number. 
This result is obtained under the approximation that $g \gg \nu^2k^3$, i.e., viscous effects are sufficiently small with respect to the gravitational effects.
Moreover, we assume that the initial perturbation is sufficiently small compared to the mean height. 
Surface tension is not considered.
We consider an initial perturbation height $\eta_0/h_0 = 1 \%$.
The two plots presented in Figure \ref{sloshingGraph} show two different cases. 
The first, on the left, is a viscous fluid, with $\nu = 0.005$.
Hence, there is a dampening effect on the perturbation, and the amplitude of the wave reduces to 0. 
The second, on the right is a simulation without viscosity. 
In this case, there is no dampening of the wave and the oscillation amplitude remains constant. 
Both simulations show good agreement with the analytical reference solution.

\subsection{Rising bubble}\label{validBubble}
The following simulation aims at verifying the boundary condition that imposes incompressibility of a cavity fully surrounded by fluid. 
The study of rising bubbles in a fluid is well established, and many approaches exist for the simulation of such two-phase flows. 
To name a few, the volume of fluid approach \cite{hysing} and levelset methods \cite{marchandiseLevelset} have been used for these simulations. 
The use of a one-phase fluid model for the simulation of such rising bubbles may not be the most adequate choice, since the dynamics inside the bubble are being neglected. 
Nevertheless, in order to validate the incompressibility boundary condition, this can be, under certain assumptions, an interesting problem to investigate. 
Indeed, considering that the dynamics of the bubble happen over a much shorter time period than that of the fluid, it is acceptable to neglect the evolution of the fluid within the bubble. 

\begin{figure}[ht!]
    \centering
    \begin{subfigure}[b]{0.45\textwidth}
        \centering
        \includegraphics[width = \textwidth]{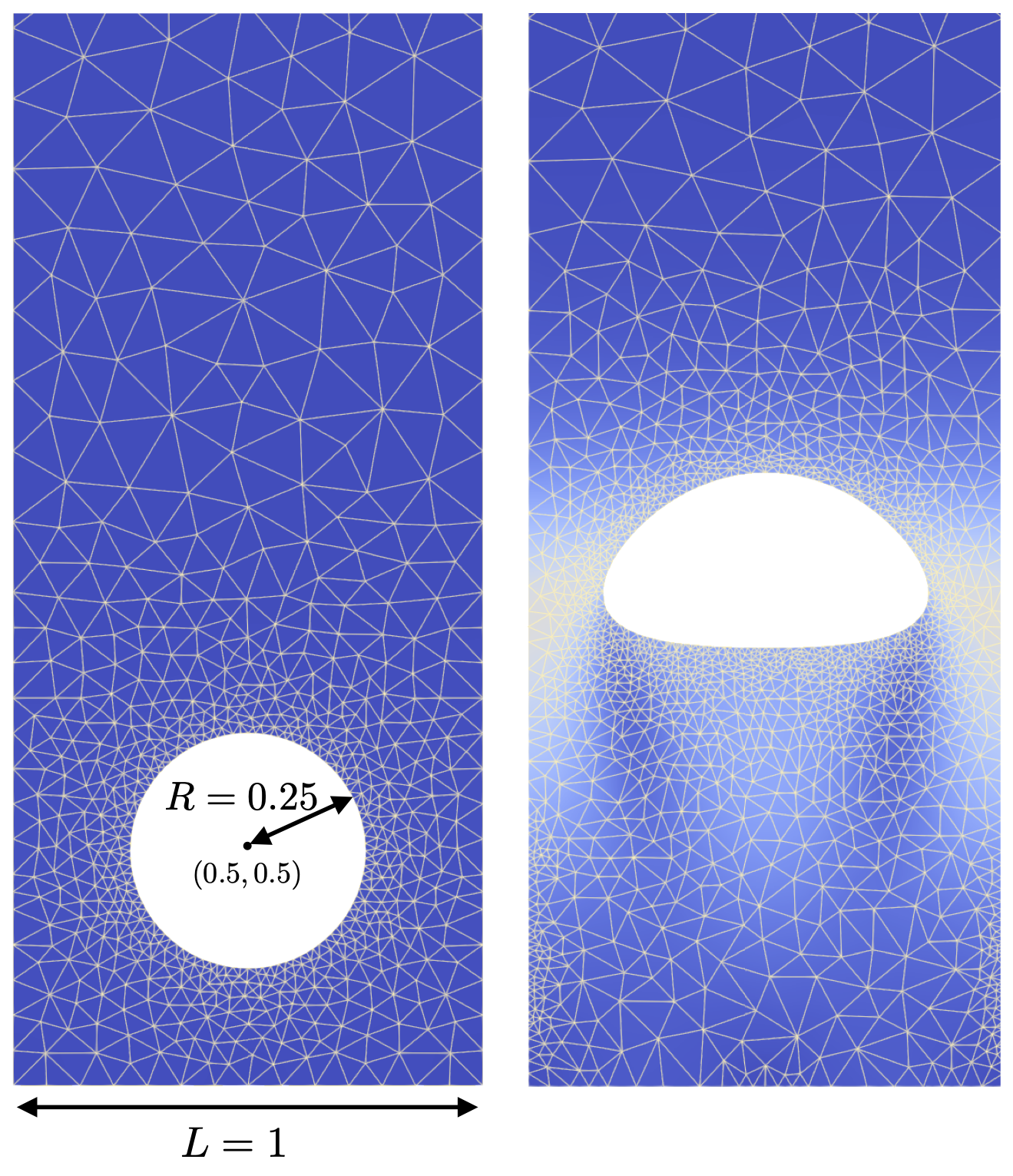}
        \caption{Evolution of the rising bubble between initial time ($t=0$s) and final time ($t=3$s).
                  The colormap corresponds to the norm of the velocity.}\label{bubbleVisu}
    \end{subfigure}
    \hfill
    \begin{subfigure}[b]{0.45\textwidth}
        \centering
        \includegraphics[width = \textwidth]{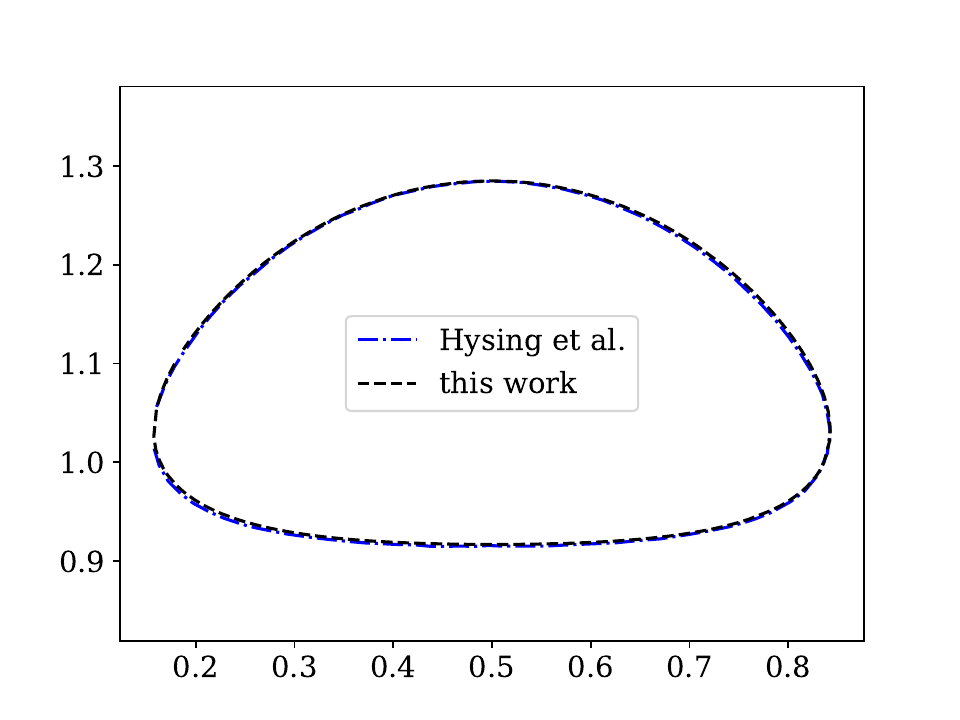}
        \vspace{.35 cm}
        \caption{Comparison between the reference solution \cite{hysing} and our PFEM approach at $t=3$s.}\label{bubbleCompar}
    \end{subfigure}
    \caption{Rising bubble experiment.}\label{bubbleAll}
\end{figure}

In the test case presented below, results have been compared with a 2-dimensionnal simulation of a rising bubble performed by Hysing et. al \cite{hysing}.
The same experimental configurations have been implemented, and the following physical parameters were considered: 
\begin{align*}
    \rho_{\text{fluid}} &= 1000 \, \si{kg.m^{-3}}\\ 
    \rho_{\text{bubble}} &=  100 \,\si{kg.m^{-3}}\\
    \mu_{\text{fluid}} &= 10 \, \si{Pa.s}\\ 
    g &= 0.98\, \si{m.s^{-2}}\\
    \sigma &= 24.5\,  \si{N.m^{-1}} 
\end{align*}
Here, $\rho_{\text{fluid}}$ and $\mu_{\text{fluid}}$ are the density and viscocity of the fluid outside the bubble, $\rho_{\text{bubble}}$ the density of the fluid inside the bubble, $g$ the gravitational acceleration and $\sigma$ the surface tension.
The non dimensional Reynolds and Eotvos numbers are defined as follows:
$$Re = \frac{\rho_{\text{fluid}} U_g L}{\mu_{\text{fluid}}} = 35, \hspace{0.5 cm} Eo = \frac{\rho_{\text{fluid}}U_g^2L}{\sigma} = 10.$$
The characteristic length scale $L = 2R$, and the gravitational velocity is given by $U_g = \sqrt{g2R}$, with $R$ the initial radius of the bubble.
The final state of the bubble is considered to be at $t=3$ s, time at which the shape of the bubble is compared. 
Figure \ref{bubbleVisu} presents the initial and final shapes of the bubble as it rises in the fluid, with the color field referring to the velocity.
Figure \ref{bubbleCompar} presents the comparison between the two-phase volume of fluid approach performed by Hysing et al., 
and the result obtained with our model, with mesh adaptation and with the boundary condition on the bubble as described in section \ref{incompbc}. 
The results confirm that the incompressibility boundary condition correctly captures the physics of the bubble, 
with only minimal differences in terms of shape compared to a full, two-phase flow simulation.

\section{Results}\label{results}
To further demonstrate the abilities of our PFEM solver, three simulations are presented in this section. 
The first is a drop of fluid falling in a bulk, followed by a dam break simulation over a dry bed, and finally a dam break over a wet bed.

\subsection{Falling drop}
\begin{figure}
    \centering
    \includegraphics[width=.6 \textwidth]{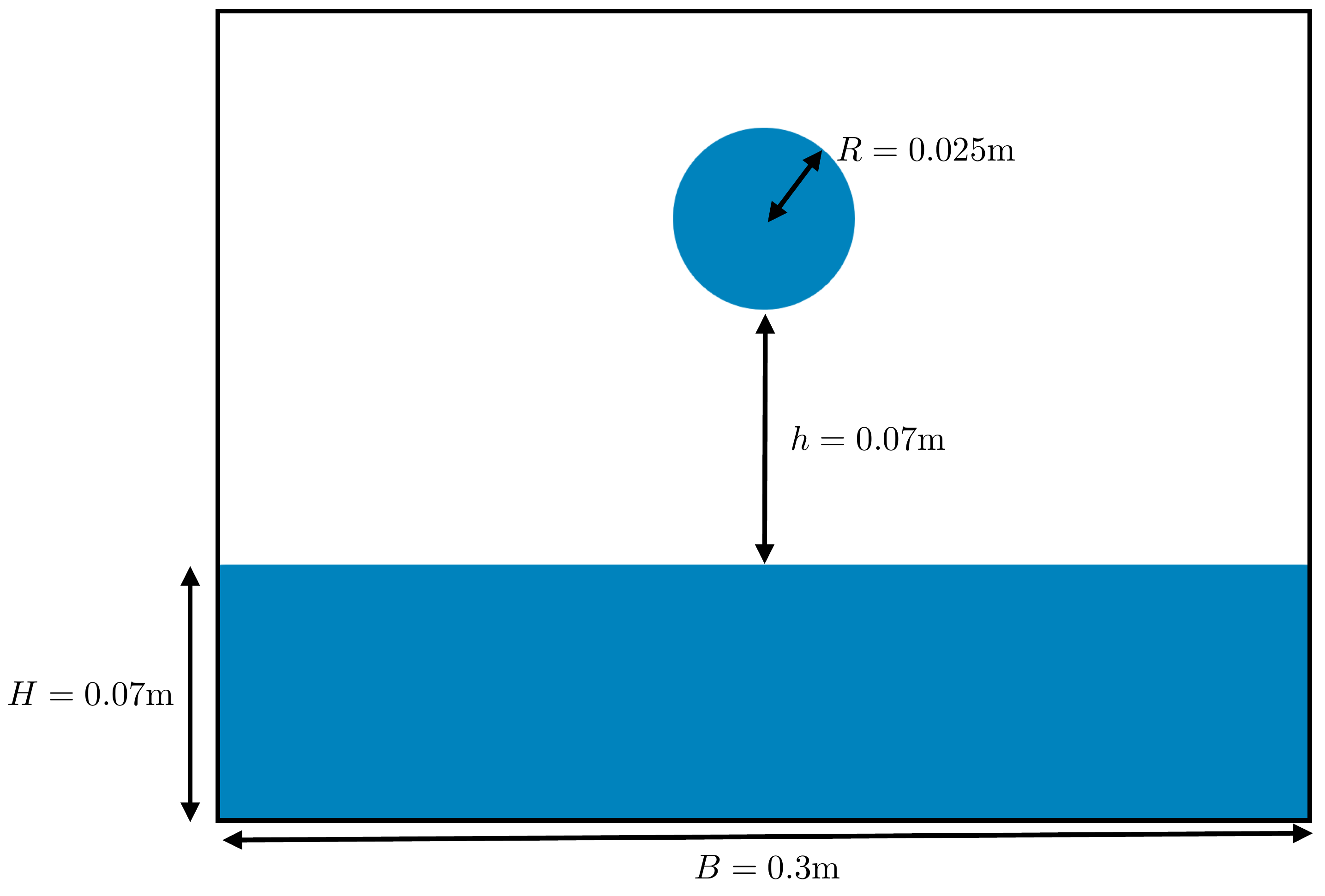}
    \caption{Initial configuration of the falling drop test case.} \label{fallingDropInitial}
\end{figure}

\begin{figure}
    \centering
    \includegraphics[width=1 \textwidth]{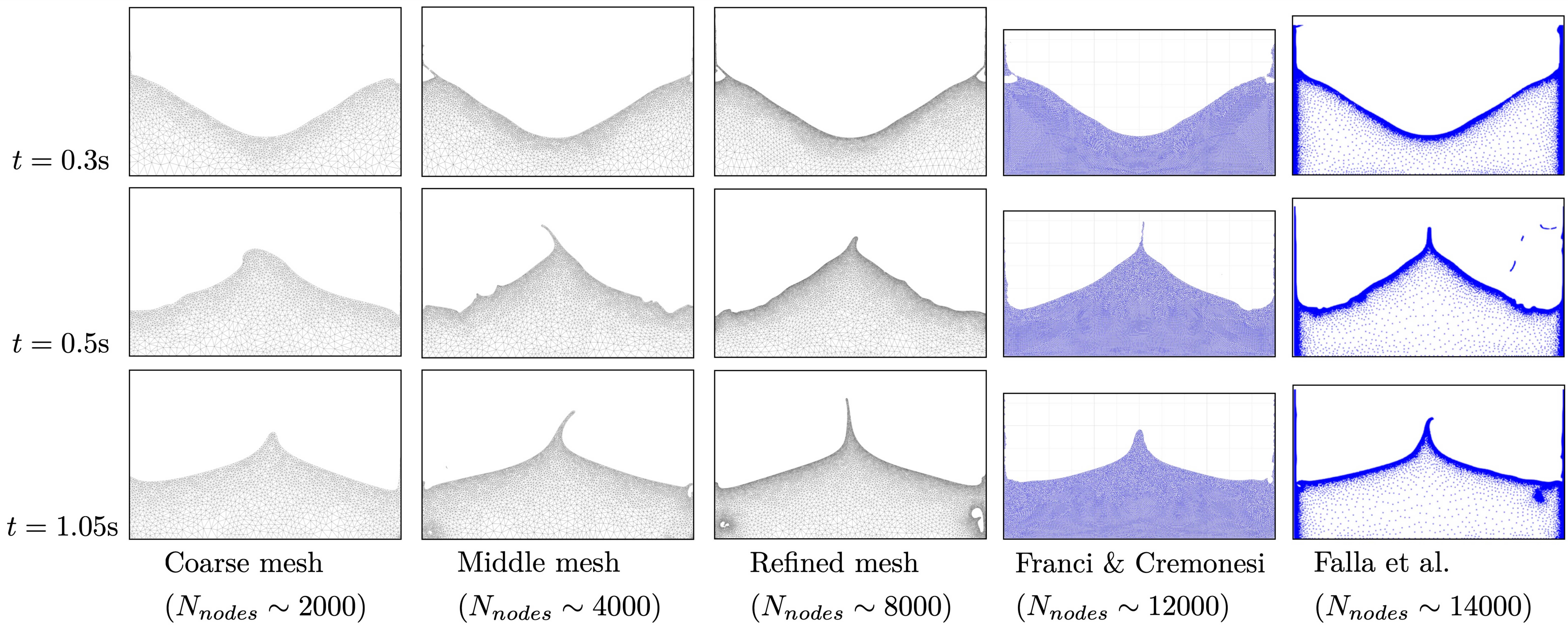}
    \caption{Falling drop simulation. Different time shots for different levels of refinement, 
    and comparison with Franci and Cremonesi \cite{franciMass} and Falla et al. \cite{falla}.} \label{dropSimu}
\end{figure}

The situation of a drop of fluid falling into a bulk is an interesting test case to study mass variation effects due to remeshing and topological changes.
Indeed, due to the impact of the drop, splashes occur and the domain changes are challenging to capture. 
Franci and Cremonesi \cite{franciMass} have studied this effect with a PFEM approach, and they were able to show that, as the mesh is refined, mass variation effects are reduced which shows the consistency of the model. 
With our proposed algorithm for mesh adaptation, the non-uniform mesh allows for more refinement near the free surface, allowing to reach higher accuracy with a smaller number of particles.
The geometry of the initial setup is presented in Figure \ref{fallingDropInitial}.
To compare with other results from the literature, the density and viscocity of the fluid are respectively 1000 \si{kg.m^{-3}} and $10^{-1}$ \si{Pa.s}.
We define a Reynolds number based on the initial diameter of the drop and the velocity as it impacts the bulk:
$$
\text{Re} = \frac{v_{\text{impact}} \rho D}{\mu} = 586.
$$
Figure \ref{dropSimu} presents a few snap shots of the falling drop simulation. 
These results are compared to simulations performed by Franci and Cremonesi \cite{franciMass} and Falla et al. \cite{falla},
who performed the same simulations with identical parameters.
The former used a uniform grid (with a number of particles estimated around 12 000), while the latter considered a locally refined mesh. 
The obtained results compare relatively well with the other results from the literature,
and similar levels of detail can be captured with smaller amounts of particles. 

\begin{figure}
    \centering
    \includegraphics[width=.8 \textwidth]{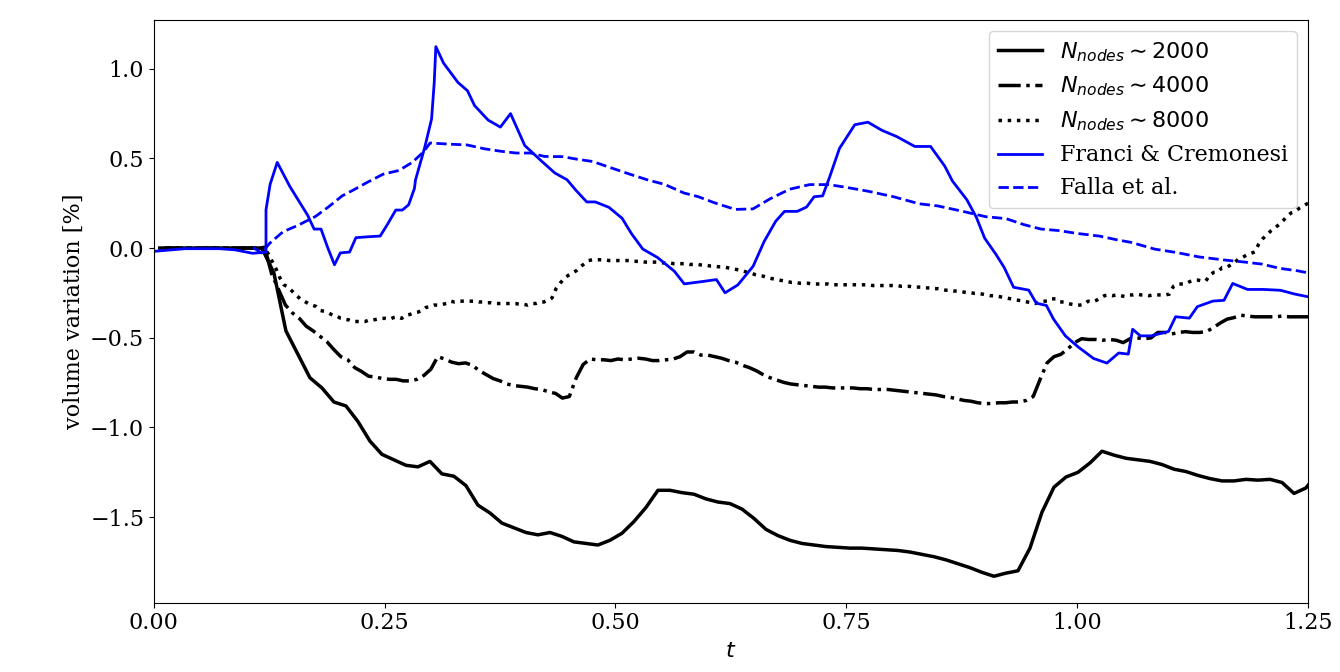}
    \caption{Volume variation over time for the falling drop simulation for different levels of refinement, and comparison with Franci and Cremonesi, and Falla et al..} \label{dropVolume}
\end{figure}

Figure \ref{dropVolume} presents the evolution of the total volume of the fluid over time. 
The variation reduces as the mesh is refined, which shows consistency of the approach.
Moreover, our most refined mesh, with a number of particles around 8000, show smaller variations compared to Franci and Cremonesi's most refined result. 
Compared to Falla et al., our approach shows smaller amounts of volume variation as well with a lesser amount of particles. 

\subsection{Dam break over a dry bed}
In the dam break simulation, a column of fluid initially at rest between three walls is set in motion due to the sudden removal of one of the side walls. 
A flow then develops along the bottom surface, until it hits an outer vertical wall. 
The flow quickly becomes turbulent as a wave collapses back onto the fluid and propagates in the other direction. 
A high amplitude sloshing regime develops, and the fluid returns to a stable state as it is slowed down by its own viscosity. 
The initial regime has been well documented, and experimental data exist to validate the numerical solvers. 

\begin{figure}
    \centering
    \includegraphics[width = .5 \textwidth]{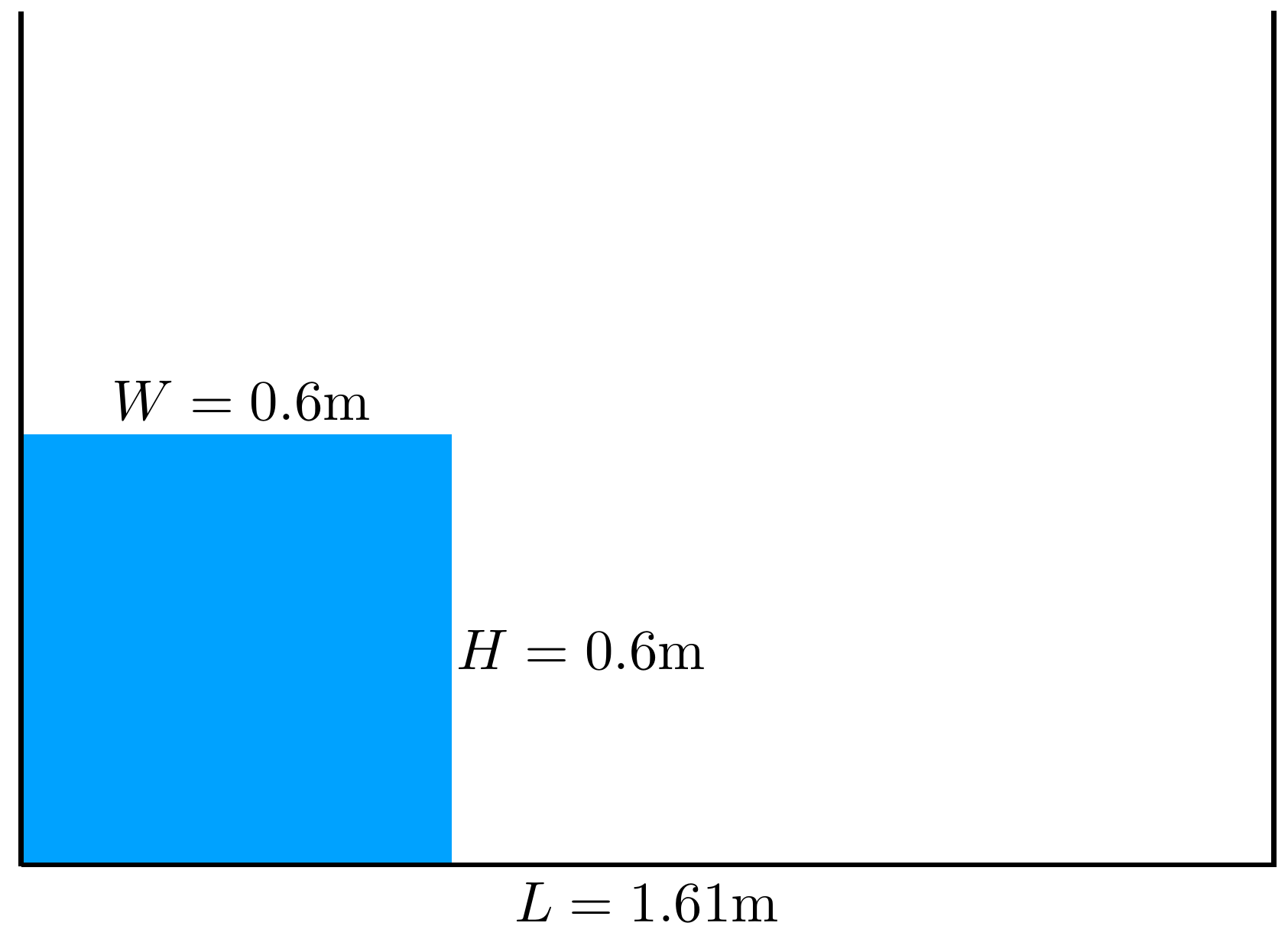}
    \caption{Geometrical configuration of the dam break experiment.}\label{dambreakConfig}
\end{figure}

\begin{figure}
    \centering
    \includegraphics[width = .8 \textwidth]{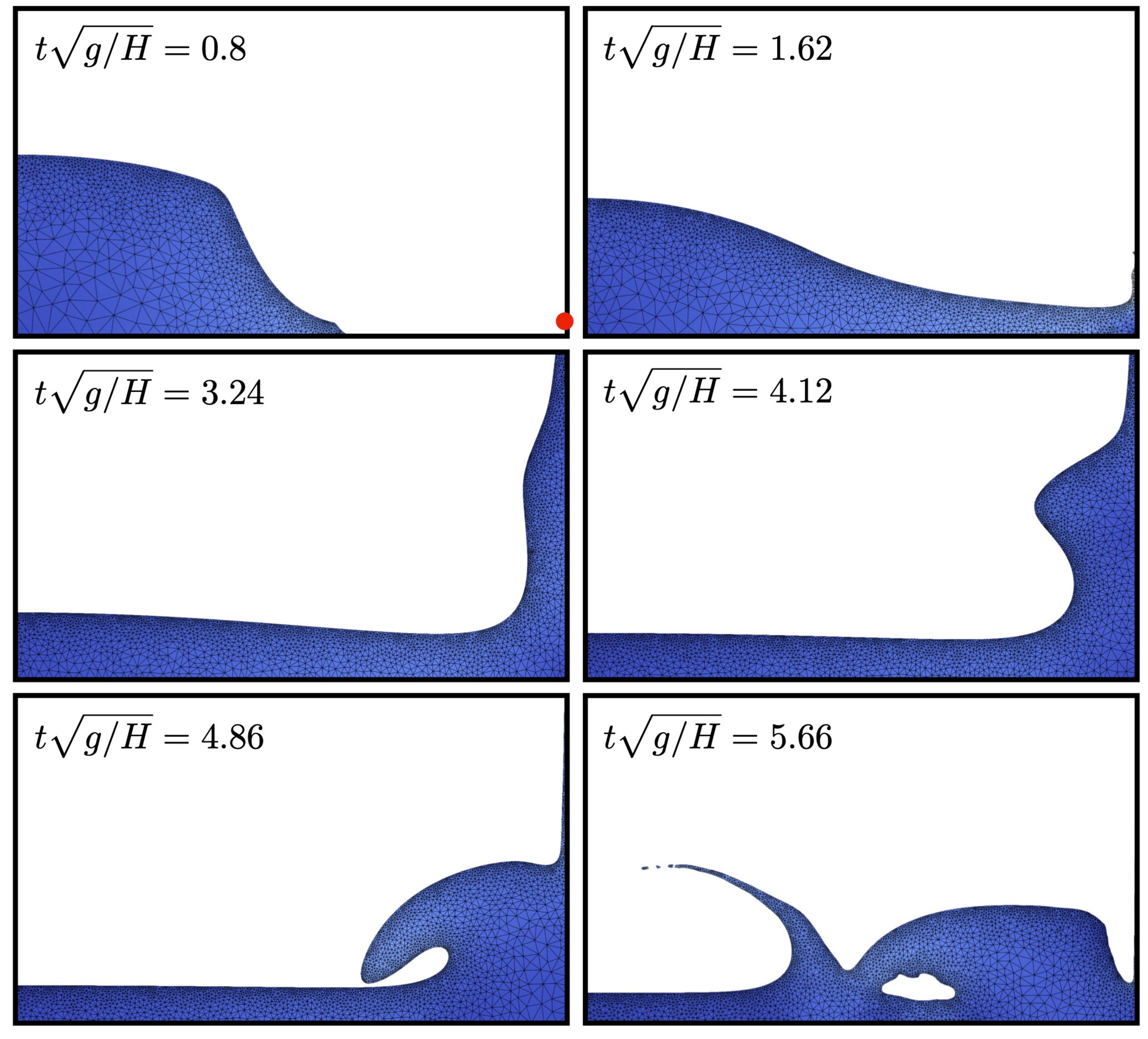}
    \caption{Different time steps of the dam break over a dry bed simulation.}\label{dambreakSteps}
\end{figure}

Non dimensional times are given by $t^* = t \sqrt{g/H}$, with $H=0.6$ \si{m} the initial height of the water column.
The fluid considered in this case is water, with density $\rho = 1000$ \si{kg.m^{-3}} and viscosity $\mu = 855 \times 10^{-6} $ \si{Pa.s}. 
The initial dimensional values of the dam break over a dry bed are summarized in Figure \ref{dambreakConfig}, and Figure \ref{dambreakSteps} presents a few time steps of the simulation.
The number of particles in this simulation is around 3000.


\begin{figure}
    \centering
    \includegraphics[width = .8 \textwidth]{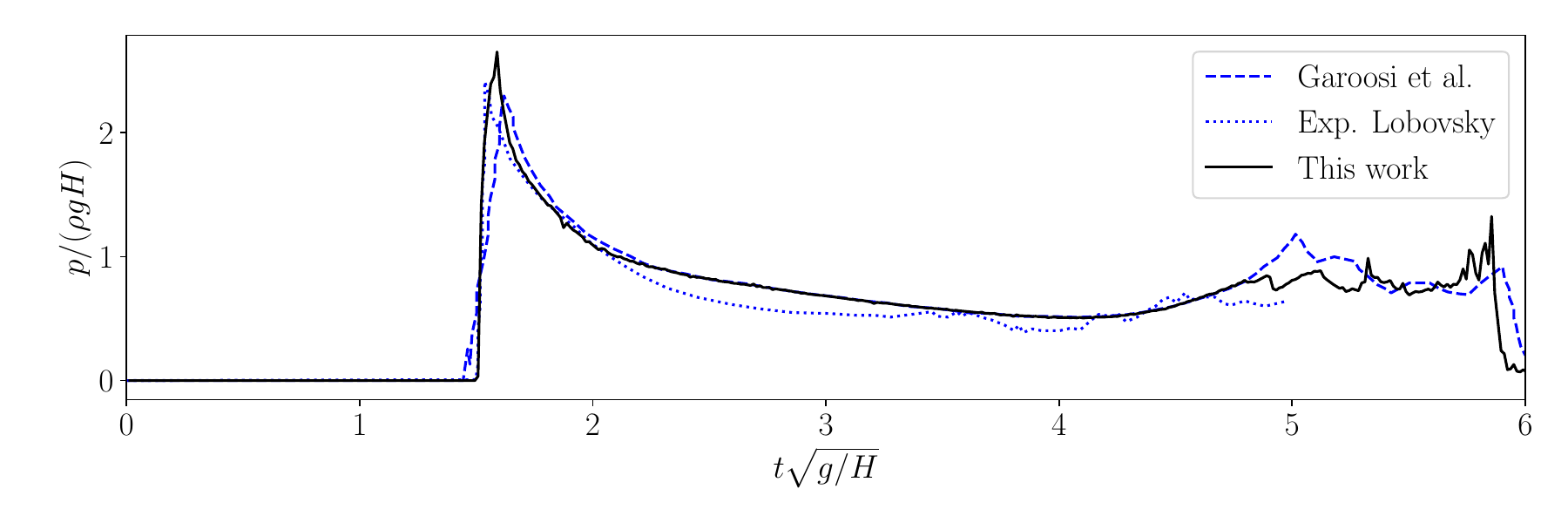}
    \caption{Pressure probed at a height of 0.03m along the right wall.}\label{pressure03}
\end{figure}

\begin{figure}
    \centering
    \includegraphics[width = .8 \textwidth]{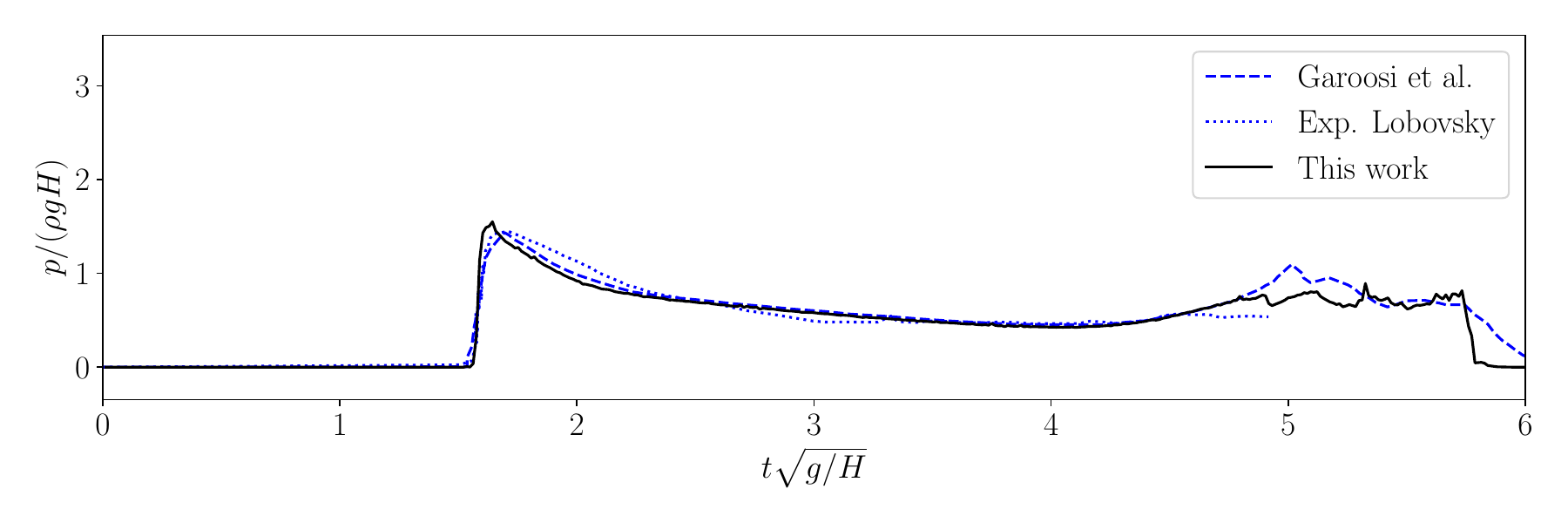}
    \caption{Pressure probed at a height of 0.08m along the right wall.}\label{pressure08}
\end{figure}

\begin{figure}
    \centering
    \includegraphics[width = .8 \textwidth]{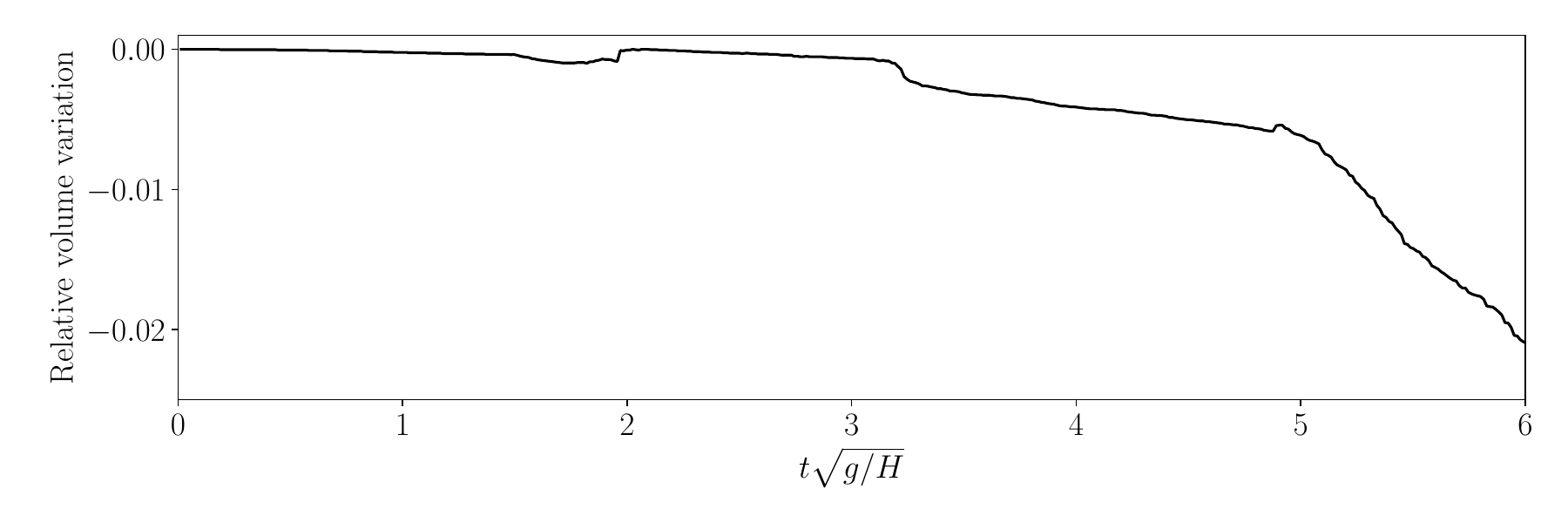}
    \caption{Relative volume variation of the fluid.}\label{volumeDambreak}
\end{figure}

To validate the obtained results, the pressure is probed on the right wall, at a height of $0.03$ m and $0.08$ m, as illustrated by the red dot on the first time step in Figure \ref{dambreakSteps}.
The evolution of the pressure is presented in figures \ref{pressure03} and \ref{pressure08}.
We compare these results with two references.
The first is a two-phase volume of fluid approach, performed by Garoosi et al. \cite{garoosiDambreak}, and the second is experimental data obtained by Lobovsky et al. \cite{lobovskyexperimental}. 
It is important to note that the experimental data available stop relatively soon after the first wave hits the right wall, but this event nevertheless matches well with our numerical simulation. 
Finally, in Figure \ref{volumeDambreak} we show that, over the course of the simulation, the relative volume variation does not exceed 2$\%$.

\subsection{Dam break over a wet bed}

This final simulation presents a water column collapse on a bed that has an initial height of water, as is usually the case on a river bed. 
The dynamics of this flow are considerably different to the dry bed case. 
Indeed, due to the presence of water downstream, the initial wave is slowed down, and a mushroom-like jet structure forms and evolves. 
This wave front then collapses onto the bed, and subsequent jets rebound on the surface and propagate, creating entrapped air cushions. 
These air cushions  are an important part of the flow, because they are responsible for a recirculating vortex to form and cause mixing within the flow. 
This again highlights the interest of the incompressibility boundary condition around bubbles presented in section \ref{incompbc}. 
This phenomenon has been experimentally observed, for example in \cite{garoosi2022,janosi}. 

\begin{figure}
    \centering
    \includegraphics[width = .95 \textwidth]{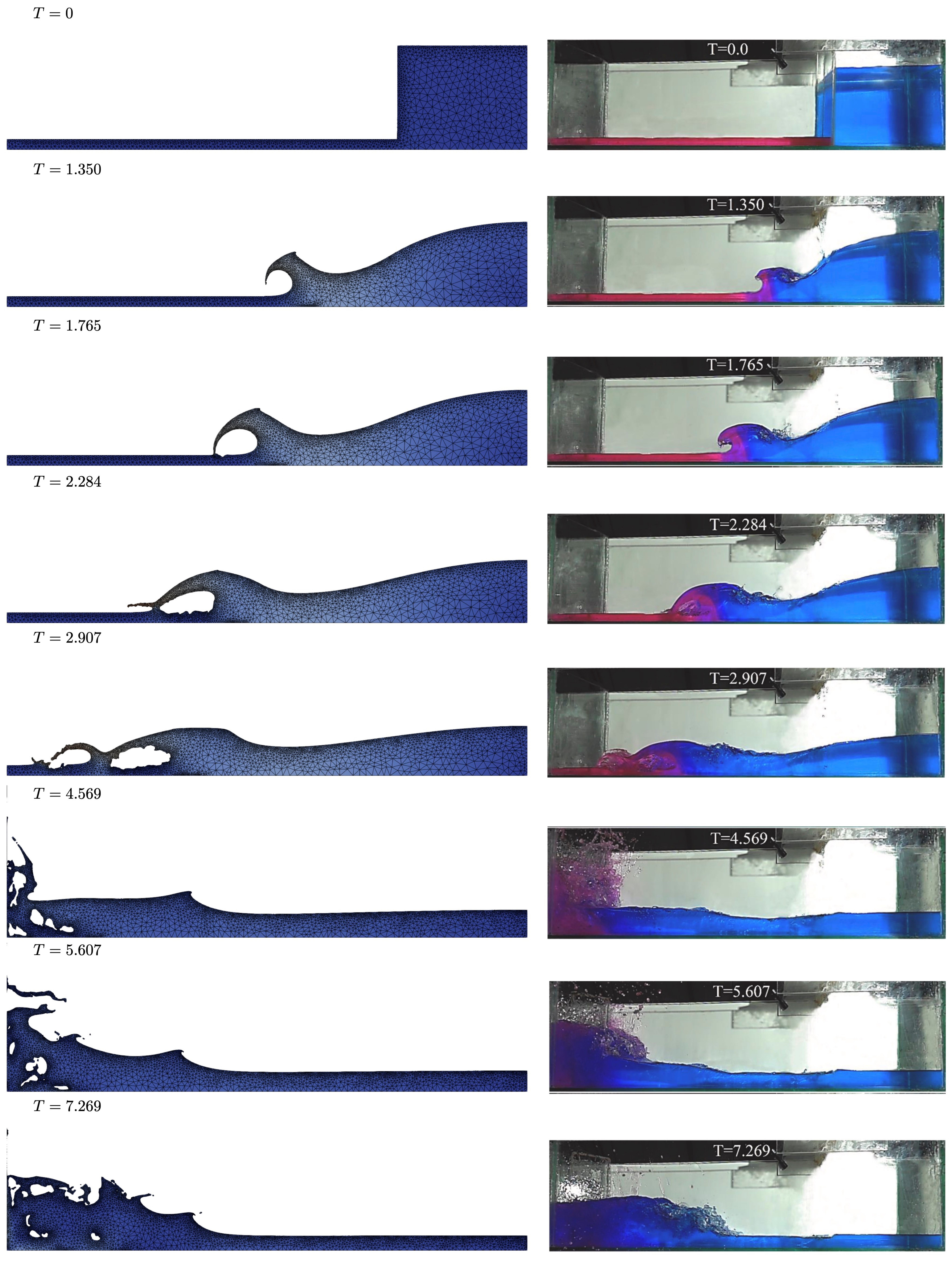}
    \caption{Different time shots of the dam break over a wet bed simulation. 
    Qualitative comparison with experiments from \cite{garoosi2022}.}\label{wetbedSimu}
\end{figure}

Figure \ref{wetbedSimu} presents a few time shots of the simulation, and are compared with experimental results from \cite{garoosi2022}.
Although the flow is in reality three-dimensional, the 2D simulation performed with the PFEM solver matches quite nicely the experimental results. 
The formation of the air cushions is clearly visible, and these bubbles remain in the flow until hitting the wall, which generates high levels of splashing. 

\section{Conclusion}\label{ccl}

This paper presented two new contributions to the particle finite element method for the simulation of free surface flows.
The first, focusing on the re-meshing step, is an algorithm that adapts the mesh with a guarantee of quality improvement for better representation of domain geometry changes. 
The second is a boundary condition around cavities within the fluid for the simulation of bubbly flows in a single-phase flow setting.

In the present formulation of the PFEM, due to constant remeshing of the particles, the shape of the domain must be redefined at each time step. 
The $\alpha$-shape of a set of points, commonly used in PFEM solvers, can by itself lead to domain detection errors.
Hence, to ensure continuity of the shape of the fluid between time steps, and to avoid artificial variations of the volume, 
it is essential to adapt the mesh, and ensure high quality of the elements within the fluid domain.
The strategy we have proposed in this work for mesh adaptation, 
by inserting particles at element circumcenters, guarantees an improvement of the quality of the elements.
The algorithm, based on Chew's approach for adaptive Delaunay refinement, not only guarantees the elements within the domain meet a certain quality criterion, 
but also allows to refine the elements based on a predefined size field.
This adaptive meshing approach thus allows to have an efficient resolution of the problem, while having sufficient accuracy in the regions of interest. 

The boundary condition around cavities allows to simulate bubbly flows while only considering one fluid.
The appearance of bubbles is almost systematic in highly turbulent free surface flows, and standard one-phase approaches do not take these features of the flow into account. 
By imposing a pressure along this surface that preserves the incompressibility, the buoyancy and the surface tension of the bubble, 
we have shown that it is possible to simulate the presence of bubbles within a single-phase model. 

Simulations show good agreement with experimental, analytical and simulation results available in the literature. 
In the two dam break simulations, the first over a dry bed and the second over an initially wet bed, 
our PFEM approach accurately captures the main features of these particularly transient free surface flows. 

Future prospects include an update to the size field, by including, for instance, solution-based error estimations. 
This will allow to better capture the main features of the flows by refining the solution in regions that include high velocity gradients.
Furthermore, analysing the sensitivity of numerical parameters such as the choice of the $\alpha$-shape limit value will add more consistency to the simulations. 
Finally, the main topic of interest will be the generalisation of the mesh adaptation algorithm to three dimensions.  

\section{Acknowledgments}

Thomas Leyssens is a Research Fellow of the F.R.S.-FNRS.

\end{document}